\documentstyle[multicol,eqsecnum,aps,prb,epsf]{revtex}

\begin{document}

\title{Exact renormalization-group study of aperiodic Ising quantum
chains\\ and directed walks}

\author{Ferenc Igl\'oi}
\address{Research Institute for Solid State Physics, H-1525 Budapest
114,  P.O. Box 49, Hungary\\ and Institute for Theoretical Physics, Szeged
University, H-6720 Szeged, Hungary} 

\author{Lo\"\i c Turban and Dragi Karevski} 
\address{Laboratoire de Physique des Mat\'eriaux, 
Universit\'e Henri Poincar\'e (Nancy I), B.P. 239,\\ 
F-54506  Vand\oe uvre l\`es Nancy cedex, France} 

\author{Ferenc Szalma}
\address{Institute for Theoretical Physics, Szeged
University, H-6720 Szeged, Hungary} 

\date{12 February 1997}
\maketitle

\begin{abstract}
We consider the Ising model and the directed walk on two-dimensional
layered lattices and show that the two problems are inherently related:
The zero-field thermodynamical properties of the Ising model are contained
in the spectrum of the transfer matrix of the directed walk. The critical
properties of the two models are connected to the scaling behavior of the
eigenvalue spectrum of the transfer matrix which is studied exactly through
renormalization for different self-similar distributions of the couplings.
The models show very rich bulk and surface critical behaviors with
nonuniversal critical exponents, coupling-dependent anisotropic scaling,
first-order surface transition, and stretched exponential critical
correlations. It is shown that all the nonuniversal critical exponents obtained
for the aperiodic Ising models satisfy scaling relations and can be expressed
as functions of varying surface magnetic exponents.  
\end{abstract}

\vglue.8truecm

\begin{multicols}{2}
\narrowtext

\section{Introduction}
\label{in}
The study of layered Ising models (IM's) has been an active field of research
during the last decades. One may mention the pioneering works on two-dimensional
(2D) periodically\cite{periodic} and
randomly\cite{mccoy68,fisher92,mikheev94} layered lattices. Similarly the
critical behavior of directed walks (DW's) in inhomogeneous\cite{fisher86} or
random media\cite{randdw} has attracted widespread interest.

Recently, following the discovery of quasicrystals,\cite{shechtman84} on the one
hand, and the progress in molecular beam epitaxy which allows the production of
good quality multilayers, on the other hand,\cite{multilayers} there
has been a growing interest in the theoretical study of phase transitions in
quasiperiodic systems and, more generally, aperiodic systems.\cite{review} These
are deterministic but nonperiodic structures which are called quasiperiodic when
the spatial fluctuations are so weak that the Fourier spectrum is still
discrete, but point symmetry is incompatible with a periodic structure.  Such
systems may be considered as intermediates between homogeneous and random ones
and, consequently, are expected to display a rich variety of critical behaviors.

\subsection{Previously known results}
\label{inta}
Most of the early works about phase transitions on aperiodic systems were done
on quasiperiodic lattices and did not show any sign of modified critical
behavior. Among these, one may mention an approximate renormalization group
treatment of the classical IM on the Penrose
lattice\cite{godreche86} and Monte Carlo renormalization group studies of
the same problem\cite{okabe88}.\cite{okabe90} Universal behavior was also
obtained in Monte Carlo simulations of the percolation problem on the Penrose
lattice and its dual\cite{sakamoto89} as well as for the statistics of
self-avoiding walks.\cite{langie92}
One may notice, as an exception, the analytical renormalization group study of
interfacial roughness, still on the Penrose lattice, where the fluctuating
interface ``feels" a Fibonacci quasiperiodic potential. In this case, a marginal
behavior was obtained for the decay of the transverse
correlations.\cite{henley87}

Probably the most studied system is the aperiodically layered 2D classical
IM and its quantum counterpart in the extreme anisotropic
limit,\cite{kogut79} the aperiodic Ising quantum chain in a transverse field. 

In the classical formulation, the energy of a configuration is given by 
\begin{equation}  
-\beta H=\sum_{k,l}K_1(k)\,\sigma_{k,l}\sigma_{k,l+1} 
+\sum_{k,l}K_2(k)\,\sigma_{k,l}\sigma_{k+1,l}\; , 
\label{Int.1}
\end{equation}
where the $\sigma$'s are the spin-$1/2$ Ising variables, and $K_1$
and $K_2$ are the exchange interactions in the vertical and
horizontal directions, respectively. Their values are the same in a vertical
layer $k$ and are modulated according to some aperiodic sequence in the
horizontal direction. 

In the extreme anisotropic limit ($K_1\!\to\!\infty,K_2\!\to\!0$),
the transfer matrix between successive rows in the vertical direction can be
written as $\exp(-\tau{\cal H})$, where $\tau\!=\!2K_1^*$ is the infinitesimal
lattice spacing in the Euclidian time direction. ${\cal H}$ is the Hamiltonian
of a spin-$1/2$ quantum Ising chain in a transverse field:
\begin{equation} 
{\cal H}=-{1\over2}\sum_{k=1}^{L} h_k\,\sigma_k^z  -{1\over2}\sum_{k=1}^{L-1}
J_k\,\sigma_k^x\sigma_{k+1}^x\; ,  
\label{Int.2} 
\end{equation}
where the $\sigma_k^{x,z}$'s are Pauli spin matrices. The transverse field
$h_k$ {\bf (}such that $h_kK_1^*$ is the dual coupling $K_1^*(k)$ given by
$\exp[-2K_1^*(k)]\!=\!\tanh K_1(k)${\bf)} plays the role of the temperature. The
coupling $J_k$ is the ratio $K_2(k)/K_1^*$. In general, due to
universality, the classical and quantum systems have the same critical
properties. Only in the presence of a marginal perturbation which leads to
nonuniversal exponents are the corresponding quantum exponents obtained by
taking an appropriate limit of the classical expressions. 

Usually the aperiodic modulation is assumed to involve only the horizontal
couplings, i.e., the two-spin interaction $J_k$ and $h_k\!=\!1$. The fluctuation
around the average $\overline{J}$ at a length scale $L$ is measured by
\begin{equation}
\Delta(L)=\sum_{k=1}^L(J_k-\overline{J})\; .
\label{Int.3}  
\end{equation}
When the aperiodic couplings are generated via substitutions using an inflation
rule, this quantity behaves as $L^\omega$, where $\omega$ is
the wandering exponent of the aperiodic sequence wich is
linked to the two leading eigenvalues of a substitution
matrix.\cite{queffelec87,dumont90} 
 
For sequences with bounded fluctuations
($\omega\!<\!0$) the aperiodic perturbation does not change the Ising critical
behavior. This was shown analytically by Tracy in the case of the Fibonacci
sequence, with $\omega\!=\!-1$, for the 2D layered IM. The Onsager
logarithmic singularity of the specific heat then keeps a nonvanishing finite
amplitude.\cite{tracy88a} The same conclusion was reached for the Ising quantum
chain with generalized Fibonacci modulations of the couplings.\cite{benza90} The
low-energy spectrum of the quantum chain which, through the gap-exponent
relation on finite critical chains, gives the values of the critical exponents
was shown to be unaffected by a quasiperiodic modulation.\cite{igloi88}
Universal behavior was also obtained with the Thue-Morse sequence and its
generalizations.\cite{doria89} In this case, the quantum chain is not
quasiperiodic but the fluctations remain bounded.

For an aperiodic sequence with unbounded fluctuations ($\omega\!>\!0$),
Tracy\cite{tracy88b} noticed that the Onsager singularity is suppressed like
in the randomly layered McCoy-Wu model.\cite{mccoy68} 

The situation was later clarified by Luck\cite{luck93a} who proposed a
generalization of the Harris criterion for quenched randomness\cite{harris74}
adapted to the case of aperiodic fluctuations of the couplings (see also
Ref.~\onlinecite{luck93b}). By comparing the mean shift of the local
temperature in the 2D layered system (governed by the
wandering exponent $\omega$), at the scale of the correlation length
of the unperturbed  system, to the deviation from the critical temperature, one
obtains a crossover exponent $\phi\!=\!1\!+\!\nu(\omega\!-\!1)$. It controls the
evolution of the amplitude of the aperiodic modulation when one approaches the
critical point. For the 2D IM with $\nu\!=\!1$, the crossover exponent is equal
to the wandering exponent so that, quite generally, the aperiodic modulation
becomes a relevant perturbation and changes the Ising critical behavior when the
fluctuations are unbounded, as conjectured by Benza {\it et al}.\cite{benza90}
One must notice that the correspondence between relevant perturbations and
unbounded fluctuations holds only when $\nu\!=\!1$. The marginal behavior
obtained for the Fibonacci sequence with bounded fluctuations in the case of the
interface roughness problem\cite{henley87} follows from Luck's criterion where 
$\nu$ is now the correlation length exponent in the transverse
direction $\nu_\perp\!=\!1/2$.  Finally let us mention that for a randomly
layered system the relevance-irrelevance criterion applies with
$\omega\!=\!1/2$.

In the same work,\cite{luck93a} Luck checked the validity of his critierion for
random and aperiodic quantum Ising chains. To treat the aperiodic problem he
considers periodic approximants, i.e., a periodic quantum chain with a large unit
cell of length $L$ in which the couplings $J_k$ are distributed according to
the aperiodic sequence. He deduces the low-energy behavior of the
fermionic excitations\cite{lieb61} $\Lambda$ from a perturbation expansion in
$\Lambda$. For the unperturbed problem at criticality, the massless excitations
have a linear low-energy dispersion relation $\Lambda\!=\!{vq}$ where $v$ is the
velocity and $q$ the wave vector. On the periodic approximant a $L$-dependent
velocity $v_L$ is obtained and the properties of the aperiodic system are
governed by the limiting behavior of $v_L$ when $L$ goes to infinity. The
behavior of the singular part of the ground-state energy, corresponding to the
free energy in the 2D classical system, is linked to the low-energy excitation
spectrum and its temperature dependence can be obtained through a scaling
argument.

For sequences with bounded fluctuations ($\omega\!<\!0$), $v_L$ is bounded and
nonvanishing in the limit $L\!\to\!\infty$ so that the Onsager logarithmic
critical singularity is preserved. For unbounded fluctuations ($\omega\!>\!0$),
the typical velocity vanishes exponentially, leading to an essential singularity
for the singular part of the ground-state energy as in the case of random chains.
Finally, when the fluctuations grow on a logarithmic scale ($\omega\!=\!0$), the
typical velocity vanishes as a nonuniversal power of $L$. The perturbation is
marginal and the specific heat exponent is negative (the logarithmic
singularity is suppressed) and varies continuously with the amplitude of the
aperiodic modulation. This marginal behavior was checked numerically.

\subsection{Renormalization-group method and main results}
\label{intb}
The results obtained so far for different aperiodic
modulations in different models are all in accordance with Luck's
criterion.\cite{turban94,igloi94,berche95,igloi95,karevski95,berche96,%
igloi96a,autres}. Most of the activity in our groups was concerned with the
study of the surface and bulk critical properties of
2D aperiodically layered IM's, either using the 1D quantum formulation or working
on a triangular lattice, making use of the star-triangle relation. 

Although some relevant
perturbations were treated in Refs.~\onlinecite{turban94}
and~\onlinecite{igloi94}, we mainly considered marginal aperiodic perturbations.
The continuously varying surface magnetic exponent $x_{m_s}\!=\!\beta_s/\nu$ was
obtained analytically for different aperiodic sequences whereas the scaling
dimension of the surface energy was conjectured on the basis of finite-size
scaling studies.\cite{karevski95,berche96,igloi96a,autres}   

The marginal aperiodic models were found to display anisotropic scaling.
\cite{berche95,igloi95,berche96} The correlation length diverges
with different exponents along and through the layers with a ratio 
$z\!=\!\nu_\parallel/\nu$, giving a continuously varying anisotropy exponent.
Such a behavior was in fact implicitly contained in Luck's work\cite{luck93a}
where a power-law dependence on $L$ was found in the marginal case. Accurate
numerical calculations of the anisotropy exponent $z$ led us to propose a simple
scaling relation  between $z$ and the surface magnetic exponents on both sides
of the system, $x_{m_s}$ and $\overline{x}_{m_s}$. The anisotropic scaling of
bulk and surface properties was extensively studied in
Ref.~\onlinecite{berche96}.

In this paper we present the results of an exact renormalization-group
(RG) study of aperiodic and hierarchical Ising and DW models. The introduction of
RG techniques into the field of phase transitions and critical phenomena has
largely contributed to our understanding of the properties of the critical
state. For instance, the RG method has given a natural explanation for the
scaling hypothesis and universality. At the same time, it has provided powerful
procedures to calculate critical exponents,\cite{rg} albeit generally using some
approximation, e.g., approximate RG transformations,  expansions in a small
parameter ($\epsilon$, $1/N,\ldots$), or numerical methods.   There are few
nontrivial problems in statistical mechanics for which the RG transformation
can be worked out exactly. One may mention the IM on the triangular
lattice\cite{hilhorst79} or different physical processes on self-similar fractal
objects.\cite{giacometti91}

Here we develop exact RG solutions for a class of 2D
layered Ising and DW models. The novel feature of our approach
is that we study both problems within the framework of the same RG
transformation. It is based on a hitherto unnoticed connection between the
eigenvalue problem for fermionic excitations which enters the solution of the
IM (Ref.~\onlinecite{lieb61}) and the transfer matrix of a DW in two
dimensions. Both problems are considered on layered lattices, such that the walk
is directed along the translationally invariant direction. The solution of the
DW, which means the diagonalization of its transfer matrix (TM), provides in
principle all the necessary information to obtain the zero-field thermodynamical
properties and correlation functions of the IM. 

The critical properties of the two models are connected to the scaling behavior
of the eigenstates of the TM at different edges of the spectrum. An exact RG
study of the eigenvalue problem of the TM is performed for
different self-similar distributions of the couplings and {\it the critical
properties of the IM and the DW are governed by two different
fixed points of the same RG transformation}.

Our method is well adapted to the case of self-similar perturbations. It is
quite different from the approximate renormalization group technique recently
introduced by Fisher to treat randomly layered systems.\cite{fisher92} In this
approach, which leads to exact results in the critical domain, instead of using
the transformation to fermions, Fischer works on the Hamiltonian itself,
reducing the energy scale of the problem by a systematic elimination of the
stronger couplings 

The bulk and surface critical properties are examined for several
aperiodic and hierarchical sequences. In all of the models we studied the
aperiodicity is marginal at the Ising fixed point and induces continuously
varying critical exponents. The bulk anisotropy exponent $z$ and the correlation
length exponents $\nu$, on the one hand, and the surface energy exponent
$x_{e_s}$ and the surface magnetization exponents $x_{m_s}$ and 
$\overline{x}_{m_s}$, on the other hand, are obtained analytically. We also prove
the previously conjectured relation between the anisotropy and surface magnetic
exponents, $z\!=\!{x}_{m_s}\!+\!\overline{x}_{m_s}>1$, which holds for sequences
which modify the critical coupling. A simple scaling picture emerges in which the
surface magnetic exponents play a fundamental role: All the nonuniversal
exponents  (except the bulk magnetic one, not considered in this work) can be
expressed as functions of these two surface magnetic exponents. The surface
energy exponent is given by $x_{e_s}\!=\!{z}\!+\!2x_{m_s}$ on one side of the
chain and $\overline{x}_{e_s}\!=\!{z}\!+\!2\overline{x}_{m_s}$ on the other,
whereas the specific heat exponent is given by $\alpha\!=\!1\!-\!{z}\!<\!0$. 

With aperiodic sequences with a
vanishing density of modified couplings, which changes the critical
behavior only locally near the surface, there is no anisotropy in the bulk of the
system ($z\!=\!1$). For sufficiently strong modified couplings, the surface
remains ordered at the bulk critical point and then the first excitation alone
scales in an anomalous way, with a continuously varying power of the size of the
system. 

Finally one may mention that these marginal aperiodic Ising sytems are closer
to periodic than to randomly layered ones. The varying exponents evolve
continuously from their unperturbed values when the aperiodic modulation grows.
In particular we checked numerically that the gap is
nonvanishing in the disordered phase, i.e., that there is no trace of a Griffiths
phase,\cite{griffiths69,fisher92} as expected for systems with bounded
fluctuations.\cite{luck93a} Even with relevant aperiodic perturbations, which by
some aspects are closer to random ones,  displaying essential singularities in
the singular part of the ground-state energy\cite{luck93a} and in the surface
magnetization,\cite{igloi94} the Griffiths phase is absent
according to a recent study.\cite{igloi97} 

A short account of our results, concerning the bulk critical behavior, has been
given in a recent Letter.\cite{igloi96b}

The structure of the paper is the following. The relation between the IM and
the DW is presented in Sec.~\ref{is}, afterwards the basic properties of
aperiodic sequences are recapitulated in Sec.~\ref{s}. In the following
sections, Secs.~\ref{pd}-\ref{f}, the RG transformation is worked out for
different aperiodic and hierarchical models which were chosen in order to
illustrate the different renormalization procedures one may use, in particular
in the treatment of the surface properties. Both the bulk and surface critical
behaviors are studied. Some relations between the critical exponents of the IM
are derived in Sec.~\ref{r} and the results are discussed in Sec.~\ref{c}.
Details about the derivation of the RG equations for the specific models are
collected in the Appendixes.

\section{Relation between the Ising quantum chain and the
directed walk model}
\label{is}

Using a Jordan-Wigner transformation,\cite{jordan28} the Ising Hamiltonian
(\ref{Int.2}) can be rewritten as a quadratic form in fermion operators. It
is then diagonalized through a canonical transformation\cite{lieb61} which
gives 
\begin{equation} 
{\cal H}=\sum_{q=1}^L\Lambda_q(\eta_q^\dagger\eta_q
-{1\over2})\; ,  
\label{I.2} 
\end{equation} 
where $\eta_q^\dagger$ and $\eta_q$ are fermion creation and anihilation
operators, respectively. The fermion excitations $\Lambda_q$ are non-negative and
satisfy the set of equations 
\begin{eqnarray}
\Lambda_q\Psi_q(k)&=&-h_k\Phi_q(k)-J_k\Phi_q(k+1)\; ,\nonumber\\
\Lambda_q\Phi_q(k)&=&-J_{k-1}\Psi_q(k-1)-h_k\Psi_q(k)\; ,
\label{I.3}
\end{eqnarray} 
with the boundary conditions $J_0\!=\!{J}_L\!=\!0$. The vectors
${\bf\Phi}_q$'s and  ${\bf\Psi}_q$'s, which are related to the
coefficients of the canonical transformation, are normalized. They enter
into the expressions of correlation functions and thermodynamical
quantities.\cite{lieb61}

\begin{figure}[t]
\epsfxsize=7.6cm
\begin{center}
\vglue-4.5truecm
\hspace*{-13truemm}\mbox{\epsfbox{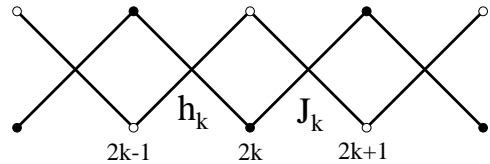}}
\end{center}
\caption{Portion of two interpenetrating diagonally layered square lattices
involved in the transfer matrix of the directed walk. The fugacities are
alternatively $h_k$ and $J_k$.}   
\label{fig1-ap-rg}  
\end{figure}

Usually one proceeds by eliminating either ${\bf\Psi}_q$ or ${\bf\Phi}_q$
in Eqs.~(\ref{I.3}) and the excitations are deduced from the solution of
one of the following eigenvalue problems:
\end{multicols}
\widetext
\noindent\rule{20.5pc}{.1mm}\rule{.1mm}{2mm}\hfill
\begin{eqnarray}
J_{k-1}h_{k-1}\Phi_q(k-1)+(J_{k-1}^2+h_k^2)\Phi_q(k)+J_kh_k\Phi_q(k+1)
&=&\Lambda_q^2\Phi_q(k)\; ,\nonumber\\
J_{k-1}h_k\Psi_q(k-1)+(J_k^2+h_k^2)\Psi_q(k)+J_kh_{k+1}\Psi_q(k+1)
&=&\Lambda_q^2\Psi_q(k)\; ,
\label{I.4}
\end{eqnarray}
\hfill\rule[-2mm]{.1mm}{2mm}\rule{20.5pc}{.1mm}
\begin{multicols}{2} 
\narrowtext
\noindent with the same boundary conditions as above. This last step can be
avoided by introducing a $2L$-dimensional vector ${\bf{V}}_q$ with
components 
\begin{equation}
V_q(2k-1)=-\Phi_q(k)\, ,\qquad V_q(2k)=\Psi_q(k)\; ,
\label{I.5}
\end{equation}
and noticing that the relations in Eqs.~(\ref{I.3}) then correspond to the
eigenvalue problem for the matrix:  
\begin{equation}
T=
\left(\begin{array}{ccccccc}
0&h_1&0&0&0&0&\cdots\\
h_1&0&J_1&0&0&0&\cdots\\
0&J_1&0&h_2&0&0&\cdots\\
0&0&h_2&0&J_2&0&\cdots\\
\vdots&\vdots&&\ddots&&\ddots&
\end{array}\right)\; .
\label{I.6}
\end{equation}
Taking the square of $T$, odd and even components of ${\bf V}_q$ are
decoupled, and one recovers the two eigenvalue equations in Eqs.~(\ref{I.4}). 
The matrix $T$ can be interpreted as the TM of a DW problem
on two interpenetrating, diagonally layered square lattices as shown in
Fig.~\ref{fig1-ap-rg}. The walker makes steps with weights $h_k$ and $J_k$
between first-neighbor sites on one of the two square lattices and the walk is
directed in the diagonal direction.

According to Eqs.~(\ref{I.3}), changing ${\bf\Phi}_q$ into $-{\bf\Phi}_q$ in
${\bf{V}}_q$, the eigenvector corresponding to $-\Lambda_q$ is obtained.
Thus all the information about the DW and the IM is contained in that part
of the spectrum with  $\Lambda_q\geq0$. Later on we shall restrict ourselves
to this sector.

Let us now consider the correlation lengths in the direction parallel to the
layers in both problems. For the DW it can be expressed as a function of
the two leading eigenvalues of the TM with
\begin{equation}
\xi_\parallel^{\rm
DW}=\left[\ln\left({\Lambda_L\over\Lambda_{L\!-\!1}}\right)\right]^{-1}
\simeq{\Lambda_L\over\Lambda_L-\Lambda_{L\!-\!1}}\; .
\label{I.7}
\end{equation}
Thus $\xi_\parallel^{\rm DW}$ is proportional to the inverse gap at the top
of the spectrum of the TM. For the IM in the disordered phase, the
correlation length is the inverse of the lowest excitation energy of the
Hamiltonian ${\cal H}$ in Eq.~(\ref{I.2}):
\begin{equation}
\xi_\parallel^{\rm IM}\sim\Lambda_1^{-1}\; .
\label{I.8}
\end{equation}
The fermionic excitation $\Lambda_1$ is also the lowest eigenvalue in the
positive sector of the TM.\cite{ordered} 

When any one of the two critical points is approached, the correlation
length of the problem diverges and the corresponding part in the TM
spectrum displays a scaling behavior. Let us consider a finite system with
transverse size $L\!\gg\!1$ and denote by $\Delta\Lambda_i$ either
$\Lambda_L\!-\!\Lambda_{L-i}$ for the DW or $\Lambda_i$ for the IM with
$i\!\ll\!{L}$. Under a change of the length scale by a factor $b\!>\!1$ such that
$L'\!=\!{L}/b$, the gaps are expected to transform as
\begin{equation}
(\Delta\Lambda_i)'=b^{y_\Lambda}\,\Delta\Lambda_i\; ,
\label{I.9}
\end{equation}
where the scaling dimension is generally different for different parts of the
spectrum. This leads to the finite-size scaling behavior
$\Delta\Lambda_i(L)\!\sim\!{L}^{-y_\Lambda}$ and, according to
Eqs.~(\ref{I.7}) and~(\ref{I.8}), the longitudinal correlation lengths
behave as $\xi_\parallel\!\sim\!{L}^{y_\Lambda}$. Since $\xi_\perp\!\sim\!{L}$,
the anisotropy exponent $z$, defined through
$\xi_\parallel\!\sim\!\xi_\perp^{z}$,\cite{aniso} is given by 
\begin{equation} 
z=y_\Lambda\; .
\label{I.10}
\end{equation}
In the case of the DW one is interested in the transverse fluctuations of
the walk which are characterized by a wandering exponent $w$ through
$\xi_\perp\!\sim\!\xi_\parallel^w$ (Refs.~\onlinecite{fisher86}
and~\onlinecite{igloi96a}) so that
\begin{equation}
w=y_\Lambda^{-1}\; .
\label{I.11}
\end{equation}

\section{Substitution matrix and relevance-irrelevance criterion}  
\protect\label{s}
In the following, except for the hierarchical
sequence, we consider sequences generated via substitutions on a finite
alphabet such that, in the case of two letters $A$ and $B$, $A\!\to\!{S(A)}$,
$B\!\to\!{S(B)}$. The properties of the sequence are governed by its substitution
matrix\cite{queffelec87,dumont90} 
\begin{equation}
M=\left(\begin{array}{cc}
n_A^{S(A)}&n_A^{S(B)}\\
n_B^{S(A)}&n_B^{S(B)}
\end{array}\right)\; ,
\label{S.1}
\end{equation}
where the matrix element $n_i^{S(j)}$ gives the number of $i$ in
$S(j)$. The matrix elements in $M^n$ give the same numbers in the
sequence obtained after $n$ iterations. When the substitution starts with
$j$, the corresponding numbers are contained in column $j$.

If ${\bf U}_\nu$ denotes the right eigenvector of $M$ with eigenvalue
$\Omega_\nu$, the asymptotic density of $i$ is given by
\begin{equation}
\rho_\infty^{(i)}={U_1(i)\over\sum_jU_1(j)}\; ,
\label{S.2}
\end{equation}
where ${\bf U}_1$ is the eigenvector corresponding to the leading eigenvalue
$\Omega_1$. The length of the sequence after $n$ iterations is related to the
leading eigenvalue through $L_n\!\sim\!\Omega_1^n$ so that $\Omega_1\!>\!1$.

In the following, each letter in the sequence is replaced by one digit or more
(for examples, $A\!=\!0$, $B\!=\!1$). Thus one obtains a sequence of digits
$f_k$  $(k\!=\!1,2,\ldots,L)$. The aperiodic Hamiltonian is defined as
in Eq.~(\ref{Int.2}) with a constant transverse field $h_k\!=h\!$ and a
modulation of the couplings following the aperiodic sequence, 
\begin{equation}
J_k=JR^{f_k}\; , 
\label{S.3}
\end{equation}
where $J$ is the unperturbed interaction and $R$ the modulation ratio.

When $f_k\!=\!0,1$, the cumulated deviation $\Delta(L)$ from the averaged
coupling $\overline{J}$ defined in Eq.~(\ref{Int.3}) scales with $L$ as
\begin{equation}
\Delta(L)=J(R-1)(n_L-L\rho_\infty)
\sim\delta\vert\Omega_2\vert^n\sim\delta L^\omega\; .
\label{S.4}  
\end{equation}
In this expression, 
\begin{equation} 
n_L=\sum_{k=1}^Lf_k\; ,\qquad\rho_\infty=\lim_{L\to\infty}{n_L\over L}
\label{S.5}
\end{equation}
give the number of digits equal to $1$ in a sequence with length $L$ and
their asymptotic density, which can be deduced from Eq.~(\ref{S.2}),
respectively; $\Omega_2$ is the next-to-leading eigenvalue of the
substitution matrix, $\delta$ measures the amplitude of the aperiodic
modulation, and $\omega$ is the wandering exponent of the sequence, given by 
\begin{equation} 
\omega={\ln\vert\Omega_2\vert\over\ln\Omega_1}\; .
\label{S.6}
\end{equation}
Thus the mean shift of the coupling strength $\overline{\delta J}(L)$ at a
length scale $L$, proportional to $L^{\omega-1}$, is governed by the
wandering exponent.

The relevance of the perturbation follows when one
compares the deviation $t$ from the critical point to the averaged temperature
shift $\overline{\delta t}\!\sim\!\overline{\delta J}(\xi)$ induced by the
aperiodicity at a length scale given by the correlation length
$\xi\!\sim\!{t}^{-\nu}$:\cite{luck93a} 
\begin{equation}
{\overline{\delta t}\over t}\sim t^{-\phi}\; ,
\qquad \phi=1+\nu(\omega-1)\; .
\label{S.7}
\end{equation}
When $\phi\!>\!0$, the ratio is divergent, which indicates a relevant
perturbation. When $\phi\!<\!0$, the
ratio vanishes at the critical point and the perturbation
is irrelevant. Finally, when $\phi\!=\!0$, the perturbation is marginal
and may lead to a nonuniversal behavior. The same conclusions can be reached
by calculating the scaling dimension of the modulation amplitude
$\delta$, which is equal to $\phi/\nu$.\cite{turban94}

For a strongly anisotropic unperturbed system $\nu$ in Eq.~(\ref{S.7}) has to
be replaced by the exponent $\nu_\perp$ of the correlation length in the
direction perpendicular to the layers in the 2D system.\cite{igloi96a}

The critical transverse field of the inhomogeneous IM 
is generally given by\cite{pfeuty79}
\begin{equation}
h_c=\lim_{L\to\infty}\prod_{k=1}^L(J_k)^{1/L}\; .
\label{S.8}
\end{equation}
Introducing the reduced coupling $\lambda\!=\!{J}/h$, its critical value on
the aperiodic quantum chain follows from Eqs.~(\ref{S.3}), (\ref{S.5}), 
and~(\ref{S.8}) as 
\begin{equation}
\lambda_c=R^{-\rho_\infty}\; .
\label{S.9}
\end{equation} 

\section{Period-doubling sequence}
\label{pd}
\subsection{Definition and general properties}
\label{pda} 
The period-doubling sequence\cite{collet80} follows from the
substitutions $A\!\to\!{S(A)}\!=\!{AB}$, $B\!\to\!{S(B)}\!=\!{AA}$. Here we make
the identification $A\!=\!0$ and $B\!=\!1$, i.e., $J_A\!=\!{J}$ and
$J_B\!=\!{JR}$ according to Eq.~(\ref{S.3}). Thus starting on $A$, after $n$
iterations, one obtains the following sequences of digits $f_k$:   
\begin{eqnarray}
n&=&0\; ,\qquad 0\nonumber\\
n&=&1\; ,\qquad 0\, 1\nonumber\\
n&=&2\; ,\qquad 0\, 1\, 0\, 0\nonumber\\
n&=&3\; ,\qquad 0\, 1\, 0\, 0\, 0\, 1\, 0\, 1\nonumber\\   
n&=&4\; ,\qquad 0\,\underline{1}\, 0\,\underline{0}\, 0\,\underline{1}\,
0\,\underline{1}\, 0\,\underline{1}\, 0\,\underline{0}\, 0\,\underline{1}\, 0\,
\underline{0}\; .
\label{PD.1}   
\end{eqnarray}

The eigenvalues of the substitution matrix are $\Omega_1\!=\!2$ and
$\Omega_2\!=\!-\!1$ so that the wandering exponent $\omega$, given
by Eq.~(\ref{S.6}), vanishes. The asymptotic density
$\rho_\infty\!=\!\rho_\infty^{(B)}\!=\!1/3$ follows from Eq.~(\ref{S.2}) and
leads to the Ising critical coupling  $\lambda_c\!=\!{R}^{-1/3}$, according to
Eq.~(\ref{S.9}).

One easily verifies on the last line of Eqs.~(\ref{PD.1}) that the $f_k$'s
satisfy the relations 
\begin{equation}
f_{2k}=1-f_k\; ,\qquad f_{2k+1}=0\; .
\label{PD.3}
\end{equation}

\subsection{Bulk critical behavior}
\label{pdb}

\begin{figure}
\epsfxsize=7.6cm
\begin{center}
\vglue-4.3truecm
\hspace*{-8.5truemm}\mbox{\epsfbox{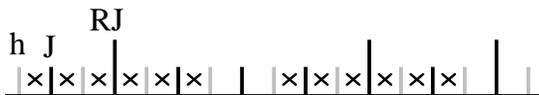}}
\end{center}
\caption{Matrix elements $T_{k,k+1}$ as a function of $k$ for the
period-doubling sequence. Components of the eigenvector to be
decimated out in the RG transformation are denoted by crosses. The heights of
solid vertical bars indicate the strength of the couplings; the grey  bars
stand for the field.}  
\label{fig2-ap-rg}  
\end{figure}

We now proceed to the exact renormalization of the eigenvalue
equations, associated with the matrix~(\ref{I.6}), which follow from
Eqs.~(\ref{I.3}) and~(\ref{I.5}). We first treat the bulk problem on a
semi-infinite system. To recover the period-doubling sequence of
interactions after one renormalization step, we eliminate triplets of
interactions $(J,RJ,J)$ indicated by crosses in Fig.~\ref{fig2-ap-rg}.

Using reduced couplings $\lambda_k\!=\!{J}_k/h$ and a reduced eigenvalue
$\widehat{\Lambda}\!=\!\Lambda/h$, the RG equations, as derived in
Appendix~\ref{apd}, are given by 
\begin{equation}
\widehat{\Lambda}'=\widehat{\Lambda}\,{c-d\over R\lambda^3}\; ,\qquad
\lambda'={c\over R\lambda^2}\; ,
\label{PD.8}
\end{equation} 
where $c$ and $d$ are defined in Eqs.~(\ref{PD.5}).

According to Eqs.~(\ref{PD.6}) and~(\ref{PD.7}), the components of the
eigenvectors transform as  
\begin{equation}
V'(2k)=V(8k)\; ,\qquad V'(2k+1)=V(8k+1)\; .
\label{PD.9}
\end{equation}

The bulk IM fixed point corresponds to
\begin{equation}
\widehat{\Lambda}^*=0\; ,\qquad \lambda^*=-R^{-1/3}\; ,
\label{PD.10}
\end{equation}
which, using Eqs.~(\ref{PD.5}), leads to
\begin{equation}
c^*=-1\; ,\qquad d^*=1+R^{2/3}+R^{-2/3}\; .
\label{PD.11}
\end{equation}
Thus the eigenvalues of the linearized transformation are given by
\begin{equation}
b^z=\left.{\partial\widehat{\Lambda}'\over\partial\widehat{\Lambda}}\right|^*
=(R^{1/3}+R^{-1/3})^2\; ,\qquad
b^{y_t}=\left.{\partial\lambda'\over\partial\lambda}\right|^*=4\; ,
\label{PD.12}
\end{equation}
and, with $b\!=\!4$, one obtains
\begin{equation}
z={\ln(R^{1/3}+R^{-1/3})\over\ln2}\; ,\qquad y_t=\nu^{-1}=1\; ,
\label{PD.13}
\end{equation}
thus confirming the conjecture of Ref.~\onlinecite{berche95}.

The top of the spectrum, which governs the behavior of the DW,
scales to a fixed point with $\widehat{\Lambda}\!\to\!\infty$ and 
$\lambda\!\to\!\infty$. Thus it is convenient to write the RG equations in terms
of the new variables $\kappa\!=\!1/\lambda$, and
$a\!=\!\widehat{\Lambda}/\lambda$ leading to
\begin{eqnarray}
&&\kappa'=\kappa^4{R\over A}\; ,\qquad a'=a\left(1-\kappa^2{B\over
A}\right)\; ,\nonumber\\ 
&&A=(a^2-R^2)(a^2-1)^2+2\kappa^2a^2\,(1-a^2)+\kappa^4a^2\; ,\nonumber\\
&&B=(a^2-R^2)(a^2-1)+\kappa^2\,(1-2a^2)+\kappa^4\; .
\label{PD.14}
\end{eqnarray}
{}From Eqs.~(\ref{PD.14}) the DW fixed point is given by 
\begin{equation}
\kappa^*=0\; ,\qquad\Delta^*=a^{*2}-R^2=0\; .
\label{PD.15}
\end{equation}

The scaling behavior at the DW fixed point is different for the homogeneous
model, with $R\!=\!1$, and for the aperiodic one, with $R\!\ne\!1$. First we
start with the homogeneous model, where the separatrix in the 
$(\Delta,\kappa)$ plane is linear at the fixed point:
$\Delta(\kappa)\!=\!\alpha^*\kappa$. According to Eqs.~(\ref{PD.14}), a point
with coordinates $\Delta\!=\!\alpha\kappa$, $\kappa\!\ll\!1$, which lies close to
the separatrix, is repelled by the fixed point as  
\begin{equation}
\kappa'={\kappa\over\alpha^3-2\alpha}\; ,\qquad\Delta'=\Delta\left[1-
2{\alpha^2-1\over\alpha^2(\alpha^2-2)}\right]\; ,
\label{PD.16}
\end{equation}
and thus $\alpha'=\Delta'/\kappa'$ is given by
\begin{equation}
\alpha'=\alpha^4-4\alpha^2+2\; .
\label{PD.17}
\end{equation}
The fixed-point value $\alpha^*\!=\!2$ determines the equation of the separatrix 
while the leading eigenvalue of the transformation, $\epsilon_1\!=\!\partial
\alpha'/\partial\alpha|^*\!=\!16$, is connected to the gap exponent through
$y_{\Lambda}\!=\!\ln\epsilon_1/\ln b\!=\!2$. Consequently, the wandering exponent
is given by 
\begin{equation}
w={1\over y_{\Lambda}}={1\over2}\qquad (R=1)\; ,
\label{PD.18}
\end{equation}
in agreement with known results.\cite{fisher86}

For the aperiodic model the separatrix has a quadratic dependence 
$\Delta(\kappa)\!=\!\beta^*\kappa^2$ when $\kappa\!\ll\!1$, with
$\beta^*\!=\!(\sqrt{2}R\!-\!2R^2)/(1\!-\!{R}^2)$, in contrast to the linear
behavior for the homogeneous model. The scaling behavior of a point with
coordinates $\Delta\!=\!\beta\kappa^2$, $\kappa\!\ll\!1$, close to the
separatrix, can be deduced from Eqs.~(\ref{PD.14}) as $\kappa'\!\sim\!\kappa^2$
and $\Delta'\!\sim\!\Delta$, thus $\Delta'/\kappa'\!\sim\!\Delta^{1/2}/\kappa$.
Consequently, at a fixed $\kappa\!=\!\kappa'\!\ll\!1$ we obtain
$\Delta'\!\sim\!\Delta^{1/2}$, which represents a strong repulsion. This type of
scaling behavior is compatible with an essential singularity in the gaps at the
top of the spectrum: 
\begin{equation}
\Delta\Lambda_i\sim\exp(-CL^{\sigma})\; ,
\label{PD.19}
\end{equation}
with $\sigma\!=\!1/2$ since the rescaling factor is $b\!=\!4$. From
Eq.~(\ref{PD.19}) the parallel correlation length of the DW is given by
$\xi_{\parallel}^{\rm DW}\!\sim\!(\Delta\Lambda_1)^{-1}\!\sim\!\exp(CL^{1/2})$,
thus the transverse fluctuations of the walk grow anomalously, on a logarithmic
scale: 
\begin{equation}
\langle[X(t)-X(0)]^2\rangle^{1/2}\sim\ln^2(t)\; .
\label{PD.20}
\end{equation}
Here $X(t)$ denotes the position of the walker at time $t$. We note that
the same asymptotic behavior is found in the Sinai model of a one-dimensional
random walk in a random environment.\cite{sinai82}

\subsection{Surface critical behavior}
\label{pdc}
We now turn to the renormalization of the surface block, looking for the
scaling behavior of the surface temperature $t_s$. In order to do so, we apply a
modified transverse field $h_1\!=\!{h}t_s$ on the first site. As shown in
Appendix~\ref{apd} the RG transformation now generates an auxiliary
variable $\theta$, in terms of which the recursion relations are given by
\begin{equation}
t_s'^2=t_s^2\,{c-d\over c-dt_s^2}\; ,\qquad
\theta'^2=\theta^2\,{c-d\over c-dt_s^2}\; ,
\label{PD.24}
\end{equation}
where $c$ and $d$ are the parameters defined in Eqs.~(\ref{PD.5}).
The auxiliary  variable $\theta$, which does not enter into the renormalization
of $t_s$, may be discarded. 

For the bulk Ising fixed point values of $c$ and $d$ given
in Eqs.~(\ref{PD.11}), the transformation of the surface temperature gives two
surface fixed points with: 
\begin{mathletters}
\label{PD.25}
\begin{eqnarray}
&&\left.{\partial t_s'^2\over\partial
t_s^2}\right|^*=(R^{1/3}+R^{-1/3})^{-2}<1\; ,\qquad t_s^{*2}=1\;
,\label{PD.25a}\\ 
&&\left.{\partial t_s'^2\over\partial
t_s^2}\right|^*=(R^{1/3}+R^{-1/3})^2>1\; ,\qquad t_s^{*2}=0\; .\label{PD.25b}
\end{eqnarray}
\end{mathletters}
Thus the attractive fixed point in the critical surface, corresponding 
to Eq.~(\ref{PD.25a}), governs the surface critical behavior and
leads to the scaling dimension of the surface temperature, 
\begin{equation}
y_{t_s}=-{\ln(R^{1/3}+R^{-1/3})\over\ln2}\; ,
\label{PD.26}
\end{equation}
in agreement with the conjecture of Ref.~\onlinecite{berche96}.
The same quantity at the repulsive fixed point, corresponding to
Eq.~(\ref{PD.25b}), is given by
\begin{equation}
\widetilde{y_{t_s}}={\ln(R^{1/3}+R^{-1/3})\over2\ln2}\; .
\label{PD.27}
\end{equation}
  
We now consider the critical behavior of the surface magnetization of the
aperiodic IM. For a semi-infinite layered system, the surface magnetization is
simply given by the first component of the normalized eigenvector
${\bf\Phi}_1$ corresponding to the lowest fermionic excitation, which vanishes
in the ordered phase:\cite{peschel84} 
\begin{equation} 
m_s=\Phi_1(1)\; ,\qquad\sum_{k=1}^\infty\Phi_1^2(k)=1\; . 
\label{PD.28}
\end{equation}
According to Eq.~(\ref{I.5}), its scaling dimension $x_{m_s}$ can be deduced from
the renormalization of the odd components of ${\bf V}$ at the Ising fixed
point which corresponds to $\Lambda^*\!=\!0$. 

Like $V(8k\!+\!7)$ in Eqs.~(\ref{PD.5}), the odd components inside each block can
be expressed as functions of $V(8k\!+\!1)$ and $V(8k\!+\!8)$ using
Eqs.~(\ref{PD.4}b)--(\ref{PD.4}g). At the fixed point, taking Eqs.~(\ref{PD.9})
into account, one obtains 
\begin{eqnarray} &&\Phi^*(4k+1)=\Phi'^*(k+1)\; ,\nonumber\\
&&\Phi^*(4k+2)=R^{1/3}\,\Phi'^*(k+1)\; ,\nonumber\\ 
&&\Phi^*(4k+3)=-R^{-1/3}\,\Phi'^*(k+1)\; ,\nonumber\\
&&\Phi^*(4k+4)=\Phi'^*(k+1)\; .  
\label{PD.29}
\end{eqnarray}
Thus the normalization of ${\bf\Phi}^*$ leads to
\begin{equation}
\sum_{k=0}^\infty\sum_{l=1}^4\!\Phi^{*2}(4k\!+\!l)
=(R^{1/3}\!+\!{}R^{-1/3})^2\sum_{k=0}^\infty\!\Phi'^{*2}(k\!+\!1)=1\, .
\label{PD.30}
\end{equation}
Near the critical point, the surface magnetization transforms as
\begin{equation}
m_s'={\Phi'^*(1)\over\sqrt{\sum_{k=0}^\infty[\Phi'^*(k+1)]^2}}
=b^{x_{m_s}}m_s\; ,
\label{PD.31}
\end{equation}
so that, using Eq.~(\ref{PD.30}), $m_s'\!=\!(R^{1/3}\!+\!R^{-1/3})\,{m}_s$ and,
with $b\!=\!4$,  
\begin{equation}
x_{m_s}={\ln(R^{1/3}+R^{-1/3})\over2\ln2}\; ,
\label{PD.32}
\end{equation}
in agreement with an analytical result for the surface
magnetization.\cite{turban94} The thermal and magnetic surface scaling
dimensions are related through
\begin{equation}
x_{m_s}=-{1\over2}y_{t_s}\; ,
\label{PD.33}
\end{equation}
a relation conjectured in Ref.~\onlinecite{berche96}.
Furthermore, comparing Eqs.~(\ref{PD.32}) and~(\ref{PD.27}), one may verify
that
\begin{equation}
x_{m_s}=\widetilde{y_{t_s}}\; .
\label{PD.34}
\end{equation}
These relations, which are generally valid for the IM, will be discussed in
Sec.~\ref{r}.

\section{Hierarchical sequence}
\label{h}
\subsection{Definition and general properties}
\label{ha}
In the generalized hierarchical sequence associated
with an integer $m\!>\!1$, the positions $k$ of the digits $f_k$ satisfy the
relation\cite{keirstead87,igloi95} 
\begin{equation}
k=m^{f_k}(ml+p)\; ,\quad l=0,1,\ldots\; ,\quad p=1,2,\ldots,m-1\; .
\label{H.1}
\end{equation}
With $m\!=\!2$, the Huberman-Kerszberg sequence is
recovered.\cite{simon61}

We recently noticed that these hierarchical sequences can be also generated
via substitution, using an alphabet with an infinite number of letters.
Let us put the letters in correspondance with the natural numbers; the
$f_k$'s then follow from $n\!\to\!{S(n)}$ with\cite{benioff97}
\begin{equation}
S(n)=\overbrace{0\, 0\, \ldots\, 0}^{m-1}\; (n+1)\; .
\label{H.2}
\end{equation}
Starting with $n\!=\!0$, repeated applications of Eq.~(\ref{H.2}) for
$m\!=\!2$ leads to the following sequence at the fourth step: 
\begin{equation}
0\, \underline{1}\, 0\, \underline{2}\, 0\, \underline{1}\, 0\,
\underline{3}\, 0\, \underline{1}\, 0\, \underline{2}\, 0\, 
\underline{1}\, 0\, \underline{4}\; .
\label{H.3}
\end{equation}
According to Eq. (\ref{S.3}), it corresponds to the interactions $J_1\!=\!{}J$,
$J_2\!=\!{}JR$, $J_3\!=\!{}J$, $J_4\!=\!{}JR^2,\ldots$ . 

One may notice that the underlined terms $f_{2k}$ give back the original
sequence with $f_k$ replaced by $f_k\!+\!1$. The same property remains true for
$f_{mk}$ with any value of $m$ so that: 
\begin{equation}
f_{mk}\!=\!{}f_k\!+\!1\, ,\quad f_{mk+1}\!=\!
f_{mk+2}\!=\!\cdots\!=\!{}f_{mk+m-1}\!=\!0\, . 
\label{H.4}
\end{equation}

The Ising critical coupling is still given by Eqs.~(\ref{S.5})
and~(\ref{S.9}) where $n_L$ can be evaluated recursively. Using Eqs.~(\ref{H.4})
for a sequence with $L\!=\!{}m^p$, we obtain
\begin{eqnarray}
n_{m^p}&=&\sum_{k=1}^{m^{p-1}}f_{mk}=n_{m^{p-1}}+m^{p-1}\nonumber\\
&=&m^{p-1}+m^{p-2}+\cdots+1={m^p-1\over m-1}\; ,
\label{H.5}
\end{eqnarray}
which leads to
\begin{equation}
\rho_\infty={1\over m-1}\; ,\qquad\lambda_c=R^{-1/(m-1)}\; .
\label{H.6}
\end{equation}

\subsection{Bulk critical behavior}
\label{hb}
\begin{figure}[t]
\epsfxsize=7cm
\begin{center}
\vglue-4truecm
\hspace*{-10.2truemm}\mbox{\epsfbox{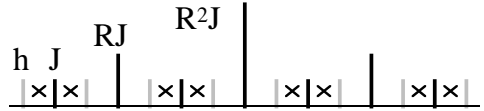}}
\end{center}
\caption{As in Fig.~\protect\ref{fig2-ap-rg} for the hierarchical
sequence with $m\!=\!2$.}  
\label{fig3-ap-rg}  
\end{figure}

In the exact renormalization group transformation, we decimate out
those sites of the lattice, which have connection by a $J$ coupling. In
such a way, blocks of $2(m\!-\!1)$ sites are eliminated as indicated by
crosses in Fig.~\ref{fig3-ap-rg}. Using the reduced variables
$\lambda\!=\!{}J/h$ and $\widehat{\Lambda}\!=\!\Lambda/h$, a lengthy calculation
detailed in Appendix~\ref{ah} leads to the transformation
\begin{equation}
\widehat{\Lambda}'=\widehat{\Lambda}\,{D_{2m-2}\over\lambda^{m-1}}+
{D_{2m-3}\over\lambda^{m-1}}\; ,\qquad\lambda^{'}=\lambda 
R\,{D_{2m-2}\over\lambda^{m-1}} \; ,
\label{H.10}
\end{equation}
where the $D$'s are determinants defined in Eq.~(\ref{H.9}).

With $m\!=\!2$ and $m\!=\!3$ one obtains
\end{multicols}
\widetext
\noindent\rule{20.5pc}{.1mm}\rule{.1mm}{2mm}\hfill
\begin{mathletters}
\label{H.11}
\begin{eqnarray}
&&\widehat{\Lambda}'={\widehat{\Lambda}\over
\lambda^2}\,(\widehat{\Lambda}^2-\lambda^2-1)\; ,\qquad\lambda'=R\,
(\widehat{\Lambda}^2-\lambda^2)\qquad(m=2)\; ,\label{H.11a}\\
&&\widehat{\Lambda}'={\widehat{\Lambda}\over
\lambda}\left[\left(\widehat{\Lambda}^2-\lambda^2
\right)^2-2\widehat{\Lambda}^2+\lambda^2+1\right]\; ,\quad\lambda'={R\over
\lambda}\left[(\widehat{\Lambda}^2-\lambda^2)^2-\widehat{\Lambda}^2\right]
\quad(m=3)\; . 
\label{H.11b} 
\end{eqnarray}
\end{mathletters}

The RG transformation in Eqs.~(\ref{H.10}) has an Ising
fixed point with $\widehat{\Lambda}^*\!=\!0$. To study the behavior of the system
close to this fixed point, we expand the determinants $D_{2m-2}$ and $D_{2m-3}$
to linear order in $\widehat{\Lambda}$:
\begin{equation} 
D_{2m-2}=(-\lambda^2)^{m-1}+O(\widehat{\Lambda}^2)\; ,\qquad
D_{2m-3}=(-1)^{m-1}\widehat{\Lambda}\,{1-\lambda^{2m-2}\over1-\lambda^2}
+O(\widehat{\Lambda}^3)\; .
\label{H.12}
\end{equation}
Putting these expressions into Eqs.~(\ref{H.10}), we obtain the location
of the Ising fixed point:
\begin{equation}  
\widehat{\Lambda}^*=0\; ,\qquad\lambda^*=(-1)^{m-1}R^{-1
/(m-1)}\; .
\label{H.13}
\end{equation}
The eigenvalues of the linearized transformation are given by
\begin{equation}
b^z=\left.{\partial\widehat{\Lambda}'\over\partial\widehat{\Lambda}}\right|^*
={|\lambda^*|^m-|\lambda^*|^{-m} \over |\lambda^*|-|\lambda^*|^{-1}}\; ,\qquad
b^{y_t}=\left.{\partial\lambda'\over\partial\lambda}\right|^*=m\; .
\label{H.14}
\end{equation}
and with $b\!=\!{}m$ one obtains 
\begin{equation}
z={\ln\left(\left|R^{m/(m-1)}-R^{-m/(m-1)}\right|\right)
-\ln\left(\left|R^{1/(m-1)}-R^{-1/(m-1)}\right|\right)
\over\ln m}\; ,\qquad y_t=\nu^{-1}=1\; ,
\label{H.15}
\end{equation}
\hfill\rule[-2mm]{.1mm}{2mm}\rule{20.5pc}{.1mm}
\begin{multicols}{2} 
\narrowtext
\noindent in agreement with the conjecture of Ref.~\onlinecite{igloi95}.

\begin{figure}
\epsfxsize=6.6cm
\begin{center}
\vglue-2.5truecm
\hspace*{-10truemm}\mbox{\epsfbox{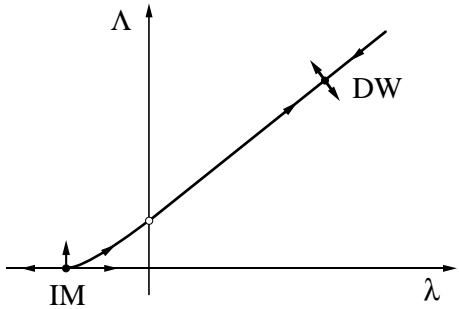}}
\vglue-.5truecm
\end{center}
\caption{Schematic RG phase diagram for the hierarchical model. There are two
nontrivial fixed points on the separatrix, at $\Lambda^*\!=\!0$ for the IM and at
a nonvanishing value of $\Lambda^*$ for the DW. The two fixed points generate
fixed lines (not shown here), parametrized by the modulation ratio $R$, in an
extended parameter space. These fixed lines govern the marginal behavior of the
IM and the DW, respectively. For the period-doubling, three-folding, and
paper-folding sequences the scaling is anomalous at the DW fixed point which is
shifted to infinity.}  
\label{fig4-ap-rg}  
\end{figure}

Another fixed point of the transformation with $\widehat{\Lambda}^*\!>\!0$
governs the critical behavior of the DW as shown in Fig.~\ref{fig4-ap-rg}. The
position of this fixed point and the values of the corresponding critical
exponents can be calculated in a closed form only for $m\!=\!2$ and $m\!=\!3$.  

The DW fixed point for $m\!=\!2$ is deduced from Eq.~(\ref{H.11a}) as
\begin{equation}
\widehat{\Lambda}^*={\sqrt{1-R+R^2}\over1-R}\; ,\qquad\lambda^*=
{R\over1-R}\; . 
\label{H.16}
\end{equation}
The leading eigenvalue of the linearized transformation is given by
\begin{equation}
\epsilon_1=R^{-1}+R+{1\over2}+
\left[\left(R^{-1}+R+{1\over2}\right)^2\!-2\right]^{1/2}\; , 
\label{H.17}
\end{equation}
and thus the wandering exponent of the walk is
\begin{equation}
w={1\over y_{\Lambda}}={\ln2\over\ln\epsilon_1}\; ,
\label{H.18}
\end{equation}
since the rescaling factor is $b\!=\!2$.

The DW fixed point for $m\!=\!3$, which follows from Eq.~(\ref{H.11b}), is
located  at: 
\begin{equation}
\widehat{\Lambda}^*={\sqrt{1+R^2}\over
1-R}\; ,\qquad\lambda^*={\sqrt{2}R\over1-R}\; .
\label{H.19}
\end{equation}
The leading eigenvalue reads
\begin{equation}
\epsilon_1=2(R^{-1}+R+1)+\left[4(R^{-1}+R+1)^2-3\right]^{1/2}\; ,
\label{H.20}
\end{equation}
and the wandering exponent of the walk is given by:
\begin{equation}
w={1\over y_{\Lambda}}={\ln3\over\ln\epsilon_1}\; ,
\label{H.21}
\end{equation}
since the rescaling factor is now $b\!=\!3$.

\subsection{Surface critical behavior}
\label{hc}
At the surface of the system we define a modified surface field $h_1\!=\!{}ht_s$
with scaling dimension $y_{t_s}$ and introduce an auxiliary variable $\theta$
to take into acount the asymmetry of the renormalized couplings. In terms of
the reduced variables, Eqs.~(\ref{H.7}) have to be supplemented for $n\!=\!0$
with the first two equations
\begin{eqnarray}
&&-\widehat{\Lambda}V(1)+\theta t_sV(2)=0\; ,\nonumber\\
&&{t_s\over\theta}V(1)-\widehat{\Lambda}V(2)+\lambda V(3)=0\; .
\label{H.22}
\end{eqnarray}
Then, besides the recursion relations of the bulk variables in
Eqs.(~\ref{H.10}), we have two more relations for the surface fields:
\begin{eqnarray}
t'_s&=&t_s\left[{D_{2m-2}+D_{2m-4} \over D_{2m-2}+t_s^2
D_{2m-4}}\right]^{1/2}\; ,\nonumber\\
\theta'&=&\theta\left[{D_{2m-2}+D_{2m-4}
\over D_{2m-2}+t_s^2 D_{2m-4}}\right]^{1/2}\; ,
\label{H.23}
\end{eqnarray}
where $D_{2m-4}$ is defined through $D_{2m-3}\!=\!\widehat{\Lambda}D_{2m-4}$.
Here, as before, the auxiliary variable $\theta$ does not enter into the
renormalization of $t_s$ and may be discarded.

As one can see from Eqs.~(\ref{H.23}), there are two surface fixed points at
$t_s^*\!=\!0$ and at $t_s^*\!=\!1$, from which the latter is stable, both on the
IM and the DW critical surfaces. Evaluating the linearized RG transformation
around the stable fixed point, one obtains, for the scaling dimension of the
surface temperature,
\begin{equation}
y_{t_s}\!=\!-{\ln \left(1+R^{2/(m-1)}+R^{4/(m-1)}+
\cdots+R^2\right)\over\ln m}\quad {\rm (IM)}
\label{H.24}
\end{equation}
for the IM as expected from numerical results\cite{igloi95} and
\begin{equation}
y_{t_s}=-{\ln R\over\ln m}\qquad {\rm (DW)}\; ,
\label{H.25}
\end{equation}
for the DW. 

The scaling dimension of the surface temperature for the DW is related to the
anomalous diffusion exponent $d_w$ on the hierarchical lattice. According to
exact results,\cite{keirstead87,giacometti91} the mean-square displacement of a
diffusive particle in a hierarchical environment is asymptotically given by
$\langle X^2(t)\rangle\!\sim\!{}t^{2/d_w}$, where 
\begin{equation}
d_w=\left\{\begin{array}{ll}
1-\ln R/\ln m\; ,\qquad &\mbox{$R<1/m$}\\
2\; ,&\mbox{$R>1/m$}
\end{array}
\right.
\label{H.26}
\end{equation}           
Thus one has $d_w\!=\!1\!+\!{}y_{t_s}$ for anomalous diffusion, i.e., with
$R\!<\!1/m$.  

One can also deduce the scaling dimension of the
surface magnetization of the IM from the rescaling of the surface component
of the eigenvector $\Phi_1(1)$. Following the same way as for the
period-doubling sequence in Sec.~\ref{pdc}, with the fixed-point
parameters in Eqs.~(\ref{H.13}), one obtains 
\begin{equation}
x_{m_s}={\ln(1+R^{2/(m-1)}+R^{4/(m-1)}+\cdots+R^2)\over2\,\ln
m}\quad{\rm (IM)}\; ,
\label{H.27}
\end{equation}
in agreement with Ref.~\onlinecite{lin95} for $m\!=\!2$ and
Ref.~\onlinecite{igloi95} for any value of $m$. One can easily check that the
scaling relations in Eqs.~(\ref{PD.33}) and~(\ref{PD.34}) are satisfied for the
hierarchical IM too.  

\section{Three-folding sequence}
\label{tf}
\subsection{Definition and general properties}
\label{tfa}
The three-folding sequence\cite{dekking83a} is generated through the
substitutions $A\!\to\!{}S(A)\!=\!{}ABA$ and $B\!\to\!{}S(B)\!=\!{}ABB$. Starting
on $A$ with $A\!=\!0$ and $B\!=\!1$, at the third iteration, one
obtains 
\begin{equation}
0\, 1\, \underline{0}\, 0\, 1\, \underline{1}\, 0\, 1\,
\underline{0} \, 0\, 1\, \underline{0}\,  0\,
1\, \underline{1}\, 0\, 1\, \underline{1}\, 0\, 1\, \underline{0}\, 
0\, 1\, \underline{1}\, 0\, 1\, \underline{0}\; . 
\label{TF.1}
\end{equation}

The substitution matrix has eigenvalues $\Omega_1\!=\!3$ and $\Omega_2\!=\!1$,
leading to the wandering exponent $\omega\!=\!0$. The asymptotic density is
$\rho_\infty\!=\!1/2$ and gives $\lambda_c\!=\!{}R^{-1/2}$ for the Ising critical
coupling. The same sequence is recovered when one keeps every third term,
underlined in Eq.~(\ref{TF.1}). The digits in between are always $0$ and $1$ so
that the following relations are obtained: 
\begin{equation}
f_{3k}=f_k\; ,\qquad f_{3k+1}=0\; ,\qquad f_{3k+2}=1\; .
\label{TF.3}
\end{equation}

\subsection{Bulk critical behavior}
\label{tfb}
\begin{figure}
\epsfxsize=7.6cm
\begin{center}
\vglue-4.8truecm
\hspace*{-8.5truemm}\mbox{\epsfbox{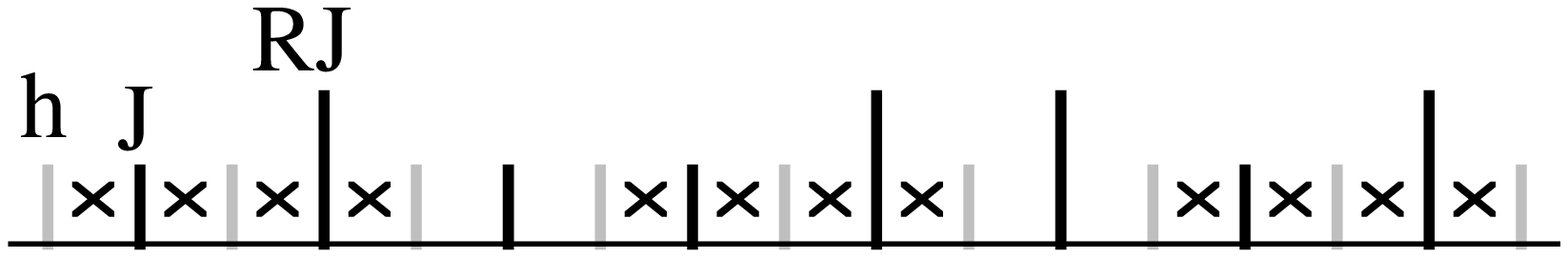}}
\end{center}
\caption{As in Fig.~\protect\ref{fig2-ap-rg} for the three-folding
sequence.}  
\label{fig5-ap-rg}  
\end{figure}

To proceed to the bulk renormalization one considers blocks of six
eigenvalue equations from which four, indicated by crosses in
Fig.~\ref{fig5-ap-rg}, are eliminated so that the rescaling factor is
now $b\!=\!3$. It is convenient to use the reduced eigenvalue
$\widetilde{\Lambda}\!=\!\Lambda/J$ and the temperaturelike parameter
$\mu\!=\!{}h/J$ as well as an auxiliary variable $\kappa$ which is needed to take
into account the form of the couplings after renormalization. The decimation
described in Appendix~\ref{atf} leads to the renormalized variables  
\begin{eqnarray}
\widetilde{\Lambda}'&=&\widetilde{\Lambda}\left[\left(1-{c\over e}\right)
\left(1-{d\over e}\right)\right]^{1/2}\, ,\nonumber\\ 
\mu'&=&{R\mu^3\over e}\,
,\qquad\kappa'^2=\kappa^2\,{c-e\over d-e}\, ,
\label{TF.8}
\end{eqnarray}
where $c$, $d$, and $e$ are the parameters defined in Eqs.~(\ref{TF.5}).
The components of the eigenvector ${\bf V}$ transform as
\begin{equation}
V'(2k)=V(6k)\; ,\qquad V'(2k+1)=V(6k+1)\; .
\label{TF.9}
\end{equation}

At the IM fixed point
\begin{equation}
\widetilde{\Lambda}^*=0\; ,\qquad\mu^*=R^{1/2}\; ,
\label{TF.10}
\end{equation}
and, according to Eqs.~(\ref{TF.5}),
\begin{equation}
c^*=-R^2\,(R+1)\, ,\quad d^*=-R\,(R+1)\, ,\quad e^*=R^2\, .
\label{TF.11}
\end{equation}
The auxiliary parameter $\kappa$, which does not enter into the
renormalization of the physical variables, need not be further considered.
The eigenvalues of the linearized RG transformation follow from
Eqs.~(\ref{TF.8}) with 
\begin{equation}
\left.{\partial\widetilde{\Lambda}'\over\partial\widetilde{\Lambda}}\right|^*
=[(2+R)(2+R^{-1})]^{1/2}\; ,\qquad
\left.{\partial\mu'\over\partial\mu}\right|^*=3\; .
\label{TF.12}
\end{equation}
Thus, with $b\!=\!3$, one obtains the bulk scaling dimensions
\begin{equation}
z={\ln[(2+R)(2+R^{-1})]\over2\ln3}\; ,\qquad y_t=\nu^{-1}=1\; ,
\label{TF.13}
\end{equation}
as expected, according to the numerical study of
Ref.~\onlinecite{berche96}. 

The top of the spectrum again scales to infinity
($\Lambda\!\to\!\infty$, $J\!\to\!\infty$) such that the DW fixed point is
located at $\widetilde{\Lambda}^*\!=\!1$, $\mu^*\!=\!0$. The equation of the
separatrix is given by $\Delta\!=\!\widetilde{\Lambda}^2\!-\!1\!=\!\gamma^*\mu$
when $\mu\!\to\!0$, i.e., close to the fixed point. The scaling behavior of a
point with coordinates $\Delta\!=\!\gamma\mu$, $\mu\!\ll\!1$ is of the form
$\Delta'\!\sim\!\Delta$ and $\mu'\!\sim\!\mu^2$, like for the period-doubling
sequence. Thus the highest gap in the TM spectrum displays an essential
singularity of the form given above in Eq.~(\ref{PD.19}) but
$\sigma\!=\!\ln2/\ln3$ since the rescaling factor is now $b\!=\!3$. The
transverse fluctuations scale similarly to Eq.~(\ref{PD.20}).

\subsection{Surface critical behavior}
\label{tfc}
At the surface we define a temperaturelike parameter $\mu_s\!=\!{}h_1/J$ with
scaling dimension $y_{t_s}$ and introduce as above an auxiliary variable
$\theta$ to take into account the asymmetry of the renormalized couplings.

A comparison of Eqs.~(\ref{TF.15}) with Eqs.~(\ref{TF.14}) leads to the
renormalized parameters 
\begin{eqnarray}
\mu_s'^2&=&\mu_s^2\,\left({\mu^2R\over e}\right)^2\,{c-e\over
c(\mu_s/\mu)^2-e}\; ,\nonumber\\
\theta'^2&=&\theta^2\,{c-e\over c(\mu_s/\mu)^2-e}\; .
\label{TF.16}
\end{eqnarray}

With the bulk Ising-fixed-point values given in Eqs.~(\ref{TF.10})
and~(\ref{TF.11}), two surface fixed points are obtained with
\begin{eqnarray}
&&\left.{\partial\mu_s'^2\over\partial\mu_s^2}\right|^*
=(2+R)^{-1}<1\; ,\qquad \mu_s^{*2}=\mu^{*2}=R\; ,\nonumber\\ 
&&\left.{\partial\mu_s'^2\over\partial\mu_s^2}\right|^*
=(2+R)>1\; ,\qquad \mu_s^{*2}=0\; .
\label{TF.17}
\end{eqnarray}
At the stable fixed point $\mu_s^{*2}\!=\!{}R$, with $b\!=\!3$, the scaling
dimension of the surface temperaturelike parameter reads
\begin{equation}
y_{t_s}=-{\ln(2+R)\over\ln3}\; ,
\label{TF.18}
\end{equation}
a result previously conjectured on the basis of a finite-size scaling
study.\cite{berche96} The same quantity at the unstable fixed point is
given by 
\begin{equation}
\widetilde{y_{t_s}}={\ln(2+R)\over2\,\ln3}\; .
\label{TF.19}
\end{equation}

The surface magnetization exponent $x_{m_s}$ follows as above from the
behavior under renormalization of the odd components $V(6k\!+\!1)$,
$V(6k\!+\!3)$, and $V(6k\!+\!5)$ which follows from
Eqs.~(\ref{TF.4}b)--(\ref{TF.4}e) and~(\ref{TF.9}). At the Ising fixed point
Eq.~(\ref{I.5}) leads to 
\begin{eqnarray}
\Phi^*(3k+1)&=&\Phi'^*(k+1)\; ,\nonumber\\
\Phi^*(3k+2)&=&-R^{1/2}\Phi'^*(k+1)\; ,\nonumber\\
\Phi^*(3k+3)&=&\Phi'^*(k+1)\; ,
\label{TF.20}
\end{eqnarray}
Making use of Eq.~(\ref{PD.30}) with $b\!=\!3$, the
scaling dimension of the surface magnetization is given by
\begin{equation}
x_{m_s}={\ln(2+R)\over2\ln3}\; ,
\label{TF.21}
\end{equation}
in agreement with a direct calculation of the surface
magnetization.\cite{berche96} Again $x_{m_s}$ satisfies the scaling
relations~(\ref{PD.33}) and~(\ref{PD.34}).  

\section{Paper-folding sequence}
\label{pf}
\subsection{Definition and general properties}
\label{pfa}
The paper-folding sequence\cite{dekking83b} results from the recurrent
folding of a sheet of paper onto itself, right over left. After unfolding, one
obtains a succession of up and down folds to which one associates a digit,
$0$ and $1$, respectively. After four steps, this process leads to the
following sequence:
\begin{equation}
0\, \underline{0}\, 1\, \underline{0}\, 0\, \underline{1}\, 1\,
\underline{0}\, 0\,  \underline{0}\, 1\, \underline{1}\, 0\, \underline{1}\, 1\; .
\label{PF.1}
\end{equation}
The sequence on the right of the central fold is the mirror image of the
left part, with each digit $f_k$ replaced by its complement $1\!-\!{}f_k$. As a
consequence, the asymptotic density is $\rho_\infty\!=\!1/2$ and the Ising
critical coupling is $\lambda_c\!=\!{}R^{-1/2}$.

The same sequence can be generated using the four letter substitutions
$A\!\to\!{S(A)}\!=\!{AC}$, $B\!\to\!{S(B)}\!=\!{DB}$, $C\!\to\!{S(C)}\!=\!{DC}$,
and $D\!\to\!{S(D)}\!=\!{AB}$ with the identification $A\!=\!00$, $B\!=\!11$,
$C\!=\!10$, and $D\!=\!01$. The leading eigenvalues of the substitution matrix,
$\Omega_1\!=\!2$ and $\Omega_2\!=\!1$, lead to a vanishing wandering exponent,
$\omega\!=\!0$.

The even terms $f_{2k}$, underlined in Eq.~(\ref{PF.1}), reproduce the
sequence itself whereas odd terms are alternatively $0$ and $1$. Thus one
has
\begin{equation}
f_{2k}=f_k\; ,\qquad f_{2k+1}={1\over2}[1+(-1)^k]\; .
\label{PF.3}
\end{equation}

\subsection{Bulk critical behavior}
\label{pfb}
\begin{figure}
\epsfxsize=7.6cm
\begin{center}
\vglue-4truecm
\hspace*{-8.5truemm}\mbox{\epsfbox{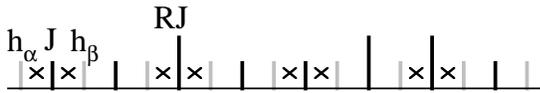}}
\end{center}
\caption{As in Fig.~\protect\ref{fig2-ap-rg} for the paper-folding
sequence.}  
\label{fig6-ap-rg}  
\end{figure}

The renormalization of the paper-folding problem is slightly more involved
than the preceding ones. In the decimation process, as shown in
Fig.~\ref{fig6-ap-rg} one eliminates blocks of two sites which
interact alternatively via $J$ or $RJ$. As a consequence,
alternating transverse fields $h_\alpha'$ and $h_\beta'$ are generated at odd and
even lattice sites, respectively. Furthermore, some auxiliary asymmetry
parameters are needed to retrieve eigenvalue equations with their original
form after renormalization. Altogether the exact RG transformation involves six
variables. 

The renormalized parameters, following from the decimation process detailed in
Appendix~\ref{apf}, are given by 
\begin{eqnarray}
&&\widetilde{\Lambda}'\!=\!\widetilde{\Lambda}\,(c_\alpha d_\beta)^{1/2}\, ,
\quad \mu_\alpha'\!=\!{\mu_\alpha\mu_\beta\over\widetilde{\Lambda}^2-1}\, ,\quad
\mu_\beta'\!=\!{\mu_\alpha\mu_\beta R\over\widetilde{\Lambda}^2-R^2}\, ,\nonumber\\
&&\kappa'^2=\kappa^2\, {c_\alpha\over d_\beta}\; ,\qquad
\kappa_\alpha'={d_\alpha\over c_\alpha}\; ,\qquad
\kappa_\beta'={c_\beta\over d_\beta}\; ,\nonumber\\
&&c_i\!=\!\kappa_i\!-\!{\mu_i^2\over\widetilde{\Lambda}^2-R^2}\, ,
\quad d_i\!=\!\kappa_i\!-\!{\mu_i^2\over\widetilde{\Lambda}^2-1}
\quad (i\!=\!\alpha,\beta).
\label{PF.9}
\end{eqnarray}

The fixed point values of interest for the IM are
\begin{eqnarray}
\widetilde{\Lambda}^*&=&0\; ,\quad\mu_\alpha^*=-R\; ,\quad
\mu_\beta^*=-1\; ,\nonumber\\
\kappa_\alpha^{*2}&=&R^2\; ,\qquad\kappa_\beta^{*2}=R^{-2}\; . 
\label{PF.10}
\end{eqnarray}
A linearization of the RG transformation, Eqs.~(\ref{PF.9}), near this fixed
point gives
\begin{eqnarray}
&&\left.{\partial\widetilde{\Lambda}'\over\partial\widetilde{\Lambda}}\right|^*
\!\!=\![(1+R)(1+R^{-1})]^{1/2}\, ,\nonumber\\
&&\left.{\partial\mu_\alpha'\over\partial\mu_\alpha}\right|^*\!\!
=\!\left.{\partial\mu_\beta'\over\partial\mu_\beta}\right|^*\!\!=\!1\, ,\ 
\left.{\partial\mu_\alpha'\over\partial\mu_\beta}\right|^*\!\!=\!{}R\, ,\ 
\left.{\partial\mu_\beta'\over\partial\mu_\alpha}\right|^*\!\!=\!{}R^{-1}\, .
\label{PF.11}
\end{eqnarray} 
The first line leads to the anisotropy exponent
\begin{equation}
z={\ln(R^{1/2}+R^{-1/2})\over\ln2}\; ,
\label{PF.12}
\end{equation}
previously conjectured in Ref.~\onlinecite{berche96},
whereas the leading eigenvalue in the linearized transformation of the
temperaturelike variables, which is equal to $2$, gives the correlation
length exponent $\nu\!=\!1$.

\subsection{Surface critical behavior}
\label{pfc}
Using Eqs.~(\ref{PF.6}) and~(\ref{PF.8}) together with Eq.~(\ref{I.5}), it may
be verified that the normalization of ${\bf\Phi}^*$ here involves two
components, $\Phi'^*(2k\!+\!1)$ and $\Phi'^*(2k\!+\!2)$. Thus the
renormalization of the surface magnetization, based on the renormalization of
the eigenvectors, becomes equivalent to a direct calculation of $m_s$. In this
case it is more convenient to introduce, besides the surface temperature, a
surface field $h_s$ conjugated to $m_s$ in the original Hamiltonian
and to study its scaling behavior. 

This can be achieved, while keeping the free-fermion
character of the Hamiltonian, through the addition of a surface term
$-\case{1}{2}h_s\sigma_0^x\sigma_1^x$ in Eq.~(\ref{Int.2}). Since there is no
transverse field acting on the first site, $\sigma_0^x$, which
commutes with ${\cal H}$, is conserved. The eigenstates of the Hamiltonian
then belong to one of the two sectors corresponding to the eigenvalues $\pm1$
of $\sigma_0^x$. Thus the  supplementary term takes the form
$\mp\case{1}{2}h_s\sigma_1^x$ and corresponds to a surface field $\pm{}h_s$
acting on $\sigma_1^x$, the sign depending on the sector.
 
The decimation of the surface block described in Appendix~\ref{apf} gives the
renormalized parameters 
\begin{equation}
\mu_s'^2=\mu_s^2\,{d_\alpha\mu_\beta^2\over d_s(\widetilde{\Lambda}^2-1)^2}\,\; ,\quad
\zeta_s'^2=\zeta_s^2\,{d_\alpha c_\beta\over d_s\kappa_\beta}\; ,\quad
\theta'^2=\theta^2\,{d_\alpha\over d_s}\; ,
\label{PF.16}
\end{equation}
where $\zeta_s\!=\!{}h_s/J$ and $\mu_s\!=\!{}h_1/J$ are reduced surface variables
whereas $\theta$ takes into account the asymmetry introduced by~$\mu_s$. 
$c_i$ and $d_i$ are the bulk parameters defined previously in
Eqs.~(\ref{PF.9}) and
$d_s\!=\!\kappa_\alpha\!-\!\mu_s^2/(\widetilde{\Lambda}^2\!-\!1)$.

Let us first consider the scaling behavior of $\mu_s$, i.e., of the
surface thermal perturbation. With the bulk values given in Eqs.~(\ref{PF.10})
one obtains two Ising surface fixed points with
\begin{eqnarray}
&&\left.{\partial\mu_s'^2\over\partial\mu_s^2}\right|^*\!\!
=(1+R)^{-1}<1\; ,\qquad \mu_s^{*2}\!=R^2\; ,\nonumber\\ 
&&\left.{\partial\mu_s'^2\over\partial\mu_s^2}\right|^*\!\!
=(1+R)>1\; ,\qquad \mu_s^{*2}=0\; .
\label{PF.17}
\end{eqnarray}
The stable fixed point corresponds to $\mu_s^{*2}\!=\!R^2$ and, with $b\!=\!2$,
the scaling dimension of the surface temperature is given by
\begin{equation}
y_{t_s}=-{\ln(1+R)\over\ln2}\; ,
\label{PF.18}
\end{equation}
as expected from numerical results,\cite{berche96} whereas
\begin{equation}
\widetilde{y_{t_s}}={\ln(1+R)\over2\ln2}\; ,
\label{PF.18a}
\end{equation}
at the unstable fixed point.

The stable fixed point values of the parameters in the equation for the
surface field variable $\zeta_s'$ lead to the transformation
\begin{equation}
\zeta_s'^2=\zeta_s^2\,(1+R^{-1})\; .
\label{PF.19}
\end{equation}
Thus, in the extended parameter space, there is a flow from $\zeta_s^{*2}\!=\!0$
to $\zeta_s^{*2}\!=\!+\infty$ and the critical behavior is governed by the fixed
point with a vanishing surface field which is unstable in the direction of
$\zeta_s^2$. Then Eq.~(\ref{PF.19}) gives the scaling dimension of the surface
field for the Ising problem, 
\begin{equation}
y_{h_s}={\ln(1+R^{-1})\over2\ln2}\; ,
\label{PF.20}
\end{equation} 
or, using Eq.~(\ref{PF.12}),
\begin{equation}
x_{m_s}=z-y_{h_s}={\ln(1+R)\over2\,\ln2}\; ,
\label{PF.21}
\end{equation}
in agreement with the scaling relations~(\ref{PD.33}) and (\ref{PD.34}) and the
analytical result of Ref.~\onlinecite{berche96}. 

\section{Fredholm sequence}
\label{f}
\subsection{Definition and general properties}
\label{fa}
The Fredholm sequence\cite{dekking83b} is obtained via substitution on
three letters, $A\!\to\!{S(A)}\!=\!{AB}$, $B\!\to\!{S(B)}\!=\!{BC}$, and
$C\!\to\!{S(C)}\!=\!{CC}$. We start the substitution process with $A$ and here,
for convenience, we number the sequence starting on $k\!=\!0$.  With $A\!=\!0$,
$B\!=\!1$, and $C\!=\!0$, after four iterations one obtains the sequence  
\begin{equation} 
\underline{0}\, 1\, \underline{1}\,
0\, \underline{1}\, 0\, \underline{0}\, 0\,  \underline{1}\, 0\,
\underline{0}\, 0\, \underline{0}\, 0\,  \underline{0}\, 0\; ,  
\label{F.1} 
\end{equation}
which is the characteristic sequence of the powers of $2$,
$f_k$ being equal to $1$ for $k\!=\!2^p$. Even underlined terms reproduce
the sequence and odd terms, except $f_1$, vanish. This gives the relations
\begin{equation}
f_{2k}=f_k\; ,\qquad f_{2k+1}=0\quad  (k>0)\; ,\qquad f_1=1\; .
\label{F.2}
\end{equation}

The leading eigenvalues of the substitution matrix are $\Omega_1\!=\!2$ and
$\Omega_2\!=\!1$, hence the wandering exponent once more vanishes. 

The number of digits equal to $1$, $n_L$, grows logarithmically  with the length
$L$, and thus the asymptotic density is $\rho_\infty\!=\!0$. The Ising critical
coupling in Eq.~(\ref{S.9}) keeps its unperturbed value $\lambda_c\!=\!1$. This
aperiodic perturbation modifies the surface critical behavior but does not
change the bulk properties, except near line defects which introduce
local marginal perturbations in the 2D IM.\cite{bariev79}   

\subsection{Bulk critical behavior}
\label{fb}
\begin{figure}
\epsfxsize=7.6cm
\begin{center}
\vglue-4truecm
\hspace*{-8truemm}\mbox{\epsfbox{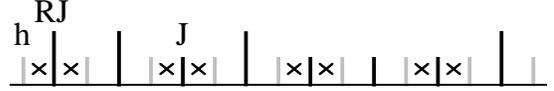}}
\end{center}
\caption{As in Fig.~\protect\ref{fig2-ap-rg} for the Fredholm
sequence.}  
\label{fig7-ap-rg}  
\end{figure}

The quantum chain is assumed to start on $k\!=\!1$, i.e., we ignore the first
digit in the sequence~(\ref{F.1}). As indicated in Fig.~\ref{fig7-ap-rg} in the
renormalization process, odd interaction terms $J_{2k+1}$ which, according
to Eqs.~(\ref{F.2}), are equal to $J$ in the bulk, are eliminated so that
$b\!=\!2$. With the same notation as before for the reduced parameters, the
renormalized variables follow from Eqs.~(\ref{F.3}) and~(\ref{F.5}) with 
\begin{mathletters}
\label{F.6}
\begin{eqnarray}
&&\widehat{\Lambda}'={\widehat{\Lambda}\over\lambda}
(\widehat{\Lambda}^2-\lambda^2-1)\; ,
\qquad\lambda'=(\widehat{\Lambda}^2-\lambda^2)\; ,\label{F.6a}\\
&&V'(2k)=V(4k)\; ,\qquad V'(2k+1)=V(4k+1)\; .\label{F.6b}
\end{eqnarray}
\end{mathletters}
Near the Ising fixed point, corresponding to
\begin{equation}
\widehat{\Lambda}^*=0\; ,\qquad\lambda^*=-1\; ,
\label{F.7}
\end{equation}
the eigenvalues of the linearized transformation
\begin{equation}
\left.{\partial\widehat{\Lambda}'\over\partial\widehat{\Lambda}}\right|^*\!\!=2\;
, \qquad\left.{\partial\lambda'\over\partial\lambda}\right|^*\!\!=2
\label{F.8}
\end{equation}
lead to the unperturbed Ising values for the anisotropy and correlation length
exponents, $z\!=\!\nu\!=\!1$.

\subsection{Surface critical behavior}
\label{fc}
With the Fredholm sequence, the decimation of the surface block
introduces a multiplicative renormalization of the first
component of the eigenvector such that (see Appendix~\ref{af}) 
\begin{equation} 
V'(1)=\theta V(1)\; ,\qquad\theta^2={(\widehat{\Lambda}^2-\lambda^2) 
(\kappa\widehat{\Lambda}^2-\lambda^2R^2-t_s^2)\over
(\widehat{\Lambda}^2-\lambda^2-1)(\kappa\widehat{\Lambda}^2-\lambda^2R^2)}\; .
\label{F.12} 
\end{equation}
The transformation of the other parameters follow from Eqs.~(\ref{F.9})
and~(\ref{F.11}) as 
\begin{eqnarray}
&&t_s'^2=t_s^2\,{(\widehat{\Lambda}^2-\lambda^2-1)(\widehat{\Lambda}^2-\lambda^2)
R^2\over(\kappa\widehat{\Lambda}^2-\lambda^2R^2)
(\kappa\widehat{\Lambda}^2-\lambda^2R^2-t_s^2)}\; ,\nonumber\\
&&\kappa'={(\widehat{\Lambda}^2-\lambda^2) 
(\kappa\widehat{\Lambda}^2-\lambda^2R^2-\kappa)\over
(\widehat{\Lambda}^2-\lambda^2-1)(\kappa\widehat{\Lambda}^2-\lambda^2R^2)}\; .
\label{F.13}
\end{eqnarray}
As before $t_s\!=\!{}h_1/h$ is the surface temperature and $\kappa$ an auxiliary
variable generated by the transformation.

{F}or the Ising fixed-point values of the bulk parameters given in
Eqs.~(\ref{F.7}) and $(\kappa\widehat{\Lambda}^2)^*\!=\!0$, one obtains two
surface fixed points with 
\begin{eqnarray} 
&&\left.{\partial t_s'^2\over\partial t_s^2}\right|^*\!\!
={2\over R^2}\; ,\qquad t_s^{*2}=0\;
,\qquad\theta^{*2}={1\over2}\; ,\nonumber\\ 
&&\left.{\partial t_s'^2\over\partial t_s^2}\right|^*\!\!
={R^2\over2}\; ,\qquad t_s^{*2}=2-R^2\; ,\qquad\theta^{*2}={1\over R^2}\; .
\label{F.14}
\end{eqnarray}

The first fixed point is stable when $R\!>\!{}R_c\!=\!\sqrt{2}$ and, since
$t_s^*\!=\!0$ corresponds to a vanishing transverse field on the first spin, the
surface is  ordered at the critical point. The second fixed point only exists
in the regime $R\!<\!{}R_c$ where it is stable. With $b\!=\!2$, the scaling
dimension of the surface temperature is given by
\begin{eqnarray}
&&y_{t_s}={1\over2}-{\ln R\over\ln2}\; ,\qquad R>\sqrt{2}\; ,\nonumber\\
&&y_{t_s}=-1+2\,{\ln R\over\ln2}\; ,\qquad R<\sqrt{2}\; .
\label{F.15}
\end{eqnarray}
These expressions were conjectured in Ref.~\onlinecite{karevski95} on the basis
of a finite-size scaling study. 

The scaling dimension of the surface magnetization
follows from the transformation of the odd components of the eigenvector at the
appropriate Ising surface fixed point. In the bulk Eqs.~(\ref{I.5}), 
(\ref{F.4}), and~(\ref{F.6b}) lead to
\begin{equation}
\Phi^*(2k+1)=\Phi^*(2k+2)=\Phi'^*(k+1)\; ,\qquad k>0\; .
\label{F.16}
\end{equation}
In the surface block, using Eqs.~(\ref{F.10}) and~(\ref{F.12}),
one obtains
\begin{equation}
\Phi^*(1)={\Phi'^*(1)\over\theta^*}\; ,\qquad\Phi^*(2)={t_s^*\Phi'^*(1)\over
\theta^*R}\; .
\label{F.17}
\end{equation}
Thus the normalization of ${\bf\Phi}^*$ gives
\begin{eqnarray}
\sum_{k=0}^\infty\sum_{l=1}^2\!\Phi^{*2}(2k+l)\!&=&\!\left(1+{t_s^{*2}\over
R^2}\right) {\Phi'^{*2}(1)\over\theta^{*2}}+\nonumber\\
&&+2\sum_{k=1}^\infty\!\Phi'^{*2}(k+1)\!=\!1\, . 
\label{F.18}
\end{eqnarray}
According to Eqs.~(\ref{F.14}), the coefficient of $\Phi'^{*2}(1)$ is equal to
$2$ at both fixed points so that
$\sum_{k=0}^\infty\Phi'^{*2}(k\!+\!1)\!=\!\case{1}{2}$. The surface magnetization
transforms according to Eq.~(\ref{PD.31}), i.e., like 
\begin{equation}
m_s'=\sqrt{2}\theta^*\, m_s\; .
\label{F.19}
\end{equation}
With $b\!=\!2$, this leads to the following scaling dimensions in the two
regimes: \begin{eqnarray}
&&x_{m_s}=0\; ,\qquad R>\sqrt{2}\; ,\nonumber\\
&&x_{m_s}={1\over2}-{\ln R\over\ln2}\; ,\qquad R<\sqrt{2}\; ,
\label{F.20}
\end{eqnarray}
as given by a direct calculation of the surface magnetization.\cite{karevski95}
The value $x_{m_s}\!=\!0$ when $R\!>\!{}R_c$ is consistent with the vanishing
surface transverse field at the fixed point. There is surface order when the
critical point is approached from the low-temperature phase and, since the
surface is one dimensional, the local magnetization vanishes discontinuously
when the bulk disorders. When $R\!<\!{}R_c$ the strength of the perturbed
couplings is not sufficient to maintain the surface order and the transition is
continuous. In this latter case the scaling relations~(\ref{PD.33})
and~(\ref{PD.34}) are still verified. 

\subsection{Aperiodic perturbation in the bulk}
\label{fd}
Let us consider the aperiodic perturbation which follows from the junction of the 
Fredholm perturbation in one half-space to its symmetric counterpart in the
other, i.e., using the symmetrized sequence
\begin{equation} 
\cdots1\,0\,0\,0\,1\,0\,1\,1\,1\,1\,0\,1\,0\,0\,0\,1\cdots
\label{F.21} 
\end{equation}
The second half of the sequence is assumed to start
on $k\!=\!1$, leaving out the term $k\!=\!0$ in the sequence~(\ref{F.1}) as for
the surface perturbation. In this way one obtains a symmetric defect in the bulk
with a vanishing asymptotic density so that $\lambda_c$ remains equal to $1$.

The simple relation between the local magnetization and the components of the
eigenvector corresponding to the lowest excitation no longer holds in the bulk
and one cannot introduce a local field term conjugated to $\sigma^x$ in the
Hamiltonian~(\ref{Int.2}) without breaking its free-fermionic character. Thus we
shall only consider the renormalization of the local temperaturelike variable
$\tau_d\!=\!{}h_1/h$. For the other parameters we keep the same notation as in
Sec.~\ref{fb} for the surface. 

\begin{figure}[t]
\epsfxsize=7.6cm
\begin{center}
\vglue-4truecm
\hspace*{-8truemm}\mbox{\epsfbox{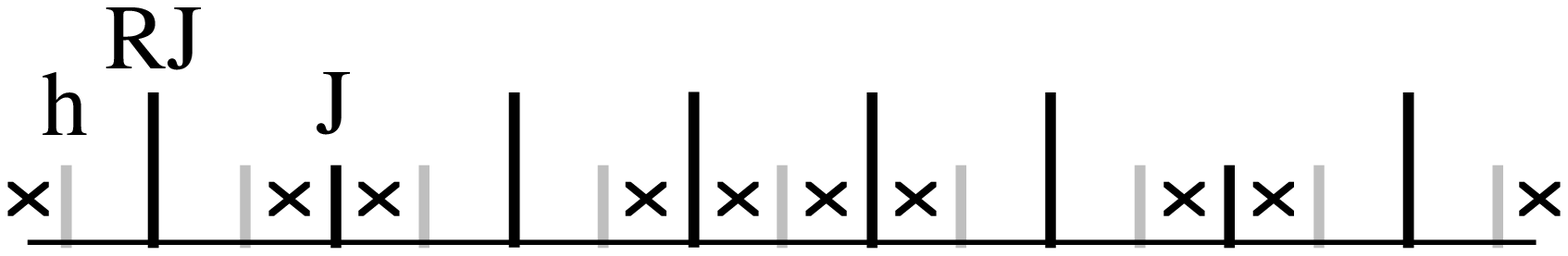}}
\end{center}
\caption{As in Fig.~\protect\ref{fig2-ap-rg} for the Fredholm defect in the
bulk.}  
\label{fig8-ap-rg}  
\end{figure}

The decimation of the central block of eigenvalue equations is illustrated in
Fig.~\ref{fig8-ap-rg}. The renormalized local variables follow from
Eqs.~(\ref{F.22}) and~(\ref{F.23}) as 
\begin{eqnarray}
&&\tau_d'=\tau_d\,{\lambda R^2(\widehat{\Lambda}^2-\lambda^2)\over
(\kappa\widehat{\Lambda}^2-\lambda^2R^2)^2-\tau_d^2\widehat{\Lambda}^2}\; ,
\nonumber\\
&&\kappa'={\widehat{\Lambda}^2-\lambda^2\over\widehat{\Lambda}^2-\lambda^2-1}\,
\left[1-{\kappa\,(\kappa\widehat{\Lambda}^2-\lambda^2R^2)-\tau_d^2\over
(\kappa\widehat{\Lambda}^2-\lambda^2R^2)^2-\tau_d^2\widehat{\Lambda}^2}\right]
\; .
\label{F.24}
\end{eqnarray}
In the critical surface, with $\widehat{\Lambda}^*\!=\!0$ and $\lambda^*\!=\!-1$
at the Ising fixed point, the RG transformation of the local variables takes the
form 
\begin{equation}
\tau_d'={\tau_d\over R^2}\; ,\qquad\kappa'={1\over2}\,
\left(1+{\kappa R^2+\tau_d^2\over R^4}\right)\; .
\label{F.25}
\end{equation}

When $R\!=\!1$, $\tau_d'\!=\!\tau_d$, which leads to a line of fixed points
parametrized by $\tau_d$, $\kappa^*\!=\!1\!+\!\tau_d$. The scaling dimension of
the local temperature vanishes as expected for a thermal line defect in the 2D
IM.\cite{bariev79} 

When $R\!\neq\!1$, two fixed points are obtained with 
\begin{eqnarray}
&&\left.{\partial \tau_d'\over\partial\tau_d}\right|^*\!\!
={1\over R^2}\; ,\qquad\tau_d^*=0\;
,\qquad\kappa^*={R^2\over2R^2-1}\; ,\nonumber\\ 
&&\left.{\partial(\tau_d^{-1})'\over\partial(\tau_d^{-1})}\right|^*\!\!
=R^2\; ,\qquad\tau_d^{*-1}=0\; ,\qquad\kappa^{*-1}=0\; .
\label{F.26}
\end{eqnarray}

In the critical surface, the fixed point at $\tau_d^*\!=\!0$ is stable when
$R\!>\!1$. The transverse field at the center of the defect vanishes at this
fixed point. Thus the defect is ordered at the critical point, like for the
surface, but the critical value of $R$ is now $R_c\!=\!1$ instead of $\sqrt{2}$.
There is no need to compensate for missing bonds as it is the case at a surface. 
When $R\!<\!1$, the second fixed point at $\tau_d^*\!=\!+\infty$ becomes stable
and leads to a second-order transition at the defect. The appropriate scaling
field $t_d$, associated with the defect temperature, which vanishes at the fixed
point is now $\tau_d^{-1}$. Hence, with $b\!=\!2$, Eqs.~(\ref{F.26}) give the
following scaling dimensions in the regimes of first- and second-order local
transition, respectively: 
\begin{eqnarray}
&&y_{t_d}=-2\,{\ln R\over\ln2}\; ,\qquad R>1\; ,\nonumber\\
&&y_{t_d}=2\,{\ln R\over\ln2}\; ,\qquad R<1\; .
\label{F.27}
\end{eqnarray}

\section{Relations between Ising model critical exponents}
\label{r}
Apart from the correlation length exponent $\nu\!=\!1$, all the critical
exponents obtained for the different aperiodic models are varying with the
amplitude of the modulation, and thus the critical behavior of these models is
nonuniversal. However, some kind of ``weak universality" still holds and there
are relations between critical exponents which follow from the fact that
the systems at the critical point obey anisotropic scaling.\cite{aniso}
A detailed analysis of the scaling behavior can be found in
Ref.~\onlinecite{berche96}.

One can notice other exponent relations which are specific for the
marginally aperiodic IM's. One such relation connects the scaling dimensions
of the energy and magnetization densities at the surface as 
\begin{equation} 
x_{e_s}=z-y_{t_s}=z+2x_{m_s}\; .
\label{R.1}
\end{equation}
It follows from Eq.~(\ref{PD.33}) and anisotropic scaling. It was conjectured in
Ref.~\onlinecite{berche96} on the basis of an assumption for the scaling
behavior of $\Phi(1)$ for low-lying excitations.  One can find another
relation which surprisingly connects bulk and surface quantities in the
form 
\begin{equation} 
z=x_{m_s}+\overline{x}_{m_s}\; .
\label{R.2}
\end{equation}
where $\overline{x}_{m_s}$ is the scaling dimension of the surface magnetization
on the right-hand side (RHS) of the system. Here we argue that the relation
in Eq.~(\ref{R.2}) is generally true for marginally aperiodic layered IM's.

In the following derivation, we consider the quantum Ising chain 
Hamiltonian ${\cal H}$ given in Eq.~(\ref{Int.2}) with homogeneous transverse
field $h_k\!=\!1$. The dynamical exponent of the model $z$ is related to the
scaling behavior of the lowest gap of the spectrum of the critical 
Hamiltonian in the form 
\begin{equation}
E_1-E_0=\Lambda_1\sim L^{-z}\; ,
\label{R.3}
\end{equation}
in a finite system of size $L$. 

The asymptotic size dependence of
$\Lambda_1(L)$ is calculated in the following approximation. First we 
determine the leading $k$ dependence of the eigenvectors $\Phi_1(k)$ and
$\Psi_1(k)$ from Eqs.~(\ref{I.4}) in such a way that the RHS's of the
equations are omitted. This approximation is justified, at the critical point
or in the ordered phase, by the fact that the second difference operators on the
LHS of the equations are $O(L^{-2})$ whereas $\Lambda_1^2$ on the RHS is
$O(L^{-2z})$ with $z\!>\!1$ for marginal aperiodic systems at criticality or
exponentially small in the ordered phase. In this approximation we obtain 
\begin{eqnarray} &&\Phi_1(L+1-k)\simeq\Phi_1(L)
\prod_{i=1}^{k-1}(-\lambda_{L-i})\left[1+\sum_{i=1}^{k-1}\prod_{j=1}^i
\lambda_{L-j}^{-2}\right]\, ,\nonumber\\
&&\Psi_1(k)\simeq\Psi_1(1)
\prod_{i=1}^{k-1}(-\lambda_i)\left[1+\sum_{i=1}^{k-1}\prod_{j=1}^i
\lambda_j^{-2}\right]\, .
\label{R.4}
\end{eqnarray}
Then the size dependence of $\Lambda_1$ is estimated from the linear
equations in Eqs.~(\ref{I.3}) as: 
\end{multicols}
\widetext
\noindent\rule{20.5pc}{.1mm}\rule{.1mm}{2mm}\hfill
\begin{mathletters}
\label{R.5}
\begin{eqnarray}
&&\Lambda_1(L)=-{\Psi_1(1)\over\Phi_1(1)}\simeq-{\Psi_1(1)\over\Phi_1(L)}
\prod_{i=1}^{L-1}(-\lambda_{i})^{-1}\left[1+\sum_{i=1}^{L-1}
\prod_{j=1}^i\lambda_{L-j}^{-2}\right]^{-1}\; ,\label{R.5a}\\
&&\Lambda_1(L)=-{\Phi_1(L)\over\Psi_1(L)}\simeq-{\Phi_1(L)\over\Psi_1(1)}
\prod_{i=1}^{L-1}(-\lambda_{i})^{-1}\left[1+\sum_{i=1}^{L-1}     
\prod_{j=1}^i\lambda_{j}^{-2}\right]^{-1}\; .
\label{R.5b}
\end{eqnarray}
\end{mathletters}
\begin{multicols}{2} 
\narrowtext
\noindent Multiplying both sides of Eqs.~(\ref{R.5a}) and~(\ref{R.5b}), one
arrives at the result 
\begin{equation}
\Lambda_1(L)\sim m_s(L)\,\overline{m}_s(L)
\prod_{i=1}^{L-1}\lambda_i^{-1}\; , 
\label{R.6}
\end{equation}
where the finite-size surface magnetizations on both sides of
the system are given by:\cite{peschel84}
\begin{eqnarray}
m_s(L)&=&\left[1+\sum_{i=1}^{L-1}\prod_{j=1}^i
\lambda_{j}^{-2}\right]^{-1/2}\; ,
\nonumber\\
\overline{m}_s(L)&=&\left[1+\sum_{i=1}^{L-1}
\prod_{j=1}^i\lambda_{L-j}^{-2} \right]^{-1/2}\; . 
\label{R.7}
\end{eqnarray}
The relation in Eq.~(\ref{R.6}), which connects the
asymptotic behavior of the lowest excitation energy and the finite-size
behavior of the surface magnetizations, is valid for general distribution of
the couplings, provided the lowest gap in the system goes to zero faster 
than~$1/L$.

In the following, we apply Eq.~(\ref{R.6}) to marginally aperiodic systems at
the critical point where, according to rigorous results,\cite{dumont90}
$\prod_{i=1}^{L-1}(\lambda_i)_c\!=\!O(1)$ for aperiodic perturbations leading to
a shift of the critical coupling. The finite-size surface magnetizations behave
as $m_{sc}(L)\!\sim\!{L}^{-x_{m_s}}$ and
$\overline{m}_{sc}(L)\!\sim\!{L}^{-\overline{x}_{m_s}}$, and thus, from
Eqs.~(\ref{R.6}) and~(\ref{R.3}), one obtains the scaling relation given in
Eq.~(\ref{R.2}).

The aperiodic sequences studied in this paper which change the bulk critical
behavior are of two kinds: either symmetric with $\lambda_k\!=\!\lambda_{L-k}$
(period doubling) or such that a perturbed coupling at $k$ corresponds to an
unperturbed coupling at $L\!-\!{k}$, which leads to $\rho_\infty\!=\!1/2$ and,
according to Eq.~(\ref{S.9}), $[\lambda_k(R)]_c\!=\![\lambda_{L-k}(R^{-1})]_c$
(paper folding, three folding). For symmetric sequences,
$\overline{m}_{sc}(L,R)\!=\!{m}_{sc}(L,R)$, and therefore
$\overline{x}_{m_s}\!=\!{x}_{m_s}$. Otherwise,
$\overline{m}_{sc}(L,R)\!=\!{m}_{sc}(L,R^{-1})$ and, consequently,
$\overline{x}_{m_s}(R)\!=\!{x}_{m_s}(R^{-1})$. {\it Thus the knowledge of a
single exponent $x_{m_s}(R)$ is sufficient to obtain all the varying exponents
studied in this paper}. Furthermore, for the period-doubling sequence
$x_{m_s}(R)$ is symmetric under the exchange of $R$ into $R^{-1}$ according
to Eq.~(\ref{PD.32}). It follows that for all the aperiodic sequences one may
rewrite the nonuniversal anisotropy exponent in Eq.~(\ref{R.2}) as
$z(R)\!=\!{x}_{m_s}(R)\!+\!{x}_{m_s}(R^{-1})$.   

For marginal aperiodic sequences which do not change the
bulk critical behavior, i.e., leave $z\!=\!1$ and $\lambda_c\!=\!1$, the
scaling  relation~(\ref{R.1}) does not hold. In this case 
$\prod_{i=1}^{L-1}(\lambda_i)_c\!=\!{R}^{-n_L}$ with the number of perturbed
couplings growing logarithmically with $L$. For the Fredholm sequence
$n_L\!=\!\ln L/\ln2$ so that the product of the couplings in Eq.~(\ref{R.6})
scales as $L^{-\ln R/\ln2}$. When the left surface is ordered at the critical
point, i.e., for $R>R_c\!=\!\sqrt{2}$, and the right surface is free, we have
$x_{m_s}\!=\!0$, $\overline{x}_{m_s}\!=\!1/2$ and the lowest excitation does not
scale as $L^{-1}$ like the rest of the spectrum, but with a continuously varying
exponent:  
\begin{equation}
\Lambda_1\sim L^{-1/2-\ln R/\ln2}\; .
\label{R.7a}
\end{equation}
When the surface magnetization vanishes at the critical point ($R\!<\!{R}_c$),
the $R$ dependence of $x_{m_s}$ in Eqs.~(\ref{F.20}) just compensates that
appearing in the product of the couplings and one recovers the normal $L^{-1}$
behavior for $\Lambda_1$.  

Finally, according to Eq.~(\ref{PD.34}), the scaling dimension of the surface
magnetization $x_{m_s}$ is equal to the scaling dimension of the surface
temperature $\widetilde{y_{t_s}}$ at the unstable fixed point.  This last
relation is a consequence of the self-duality of the Ising quantum
chain.\cite{kogut79} Using the dual Pauli spin matrices defined through
\begin{equation} 
\tau_k^z=\sigma_{k-1}^x\sigma_k^x\; ,\qquad\sigma_k^z=\tau_k^x\tau_{k+1}^x\; , 
\label{R.8} 
\end{equation}
the original Hamiltonian in Eq.~(\ref{Int.2}) is transformed into its dual:
\begin{equation}
\widetilde{\cal H}=-{1\over2}\sum_{k=1}^{L} h_k\,\tau_k^x\tau_{k+1}^x 
-{1\over2}\sum_{k=1}^{L-1} J_k\,\tau_{k+1}^z\; , 
\label{R.9}
\end{equation}
with a vanishing transverse field on the first spin. As already shown
at the begining of Sec.~\ref{pfc}, $\tau_1^x$ commutes with
$\widetilde{\cal H}$ and may be replaced by its eigenvalues $\pm1$. Thus in the
surface term $-\case{1}{2}h_1\tau_1^x\tau_2^x\!=\!\mp\case{1}{2}h_1\tau_2^x$,
$h_1\!=\!{h}t_s$ now plays the role of a surface field acting on~$\tau_2^x$. 

The unstable fixed point at $t_s^*\!=\!0$, with its associated scaling dimension
$\widetilde{y_{t_s}}$, governs the critical behavior of the dual surface
magnetization. In the duality transformation the couplings
$\lambda_k\!=\!{}h/J_k$ are changed into $\lambda_k^{-1}$ so that the surface
magnetizations on both sides of Eqs.~(\ref{R.7}) are exchanged. It follows that
the scaling dimension of $\langle\tau_2^x\rangle$ is $\overline{x}_{m_s}$ and
the dimension of the surface field is given by
$\widetilde{y_{t_s}}\!=\!{}z\!-\!\overline{x}_{m_s}$, which, according to
Eq.~(\ref{R.2}), leads to Eq.~(\ref{PD.34}). 

\section{Conclusion}
\label{c}
In this paper we have presented a unified statistical-mechanical description
of the IM and the DW on layered two-dimensional lattices, taking the extreme
anisotropic limit for the IM. The critical properties of the two problems were
deduced from the scaling behavior of the spectrum of the transfer matrix of the
DW, which is studied through exact RG transformations. For a given value of the
aperiodicity parameter $R$ the RG transformations have two nontrivial fixed
points, as shown in Fig.~\ref{fig4-ap-rg}.  The bottom of the spectrum scales to
the IM fixed point, which controls the critical behavior of the IM, whereas
the top of the spectrum scales to another fixed point, which describes the
critical properties of the DW.

The aperiodic sequences we considered have different effects on the critical
properties of the two models according to Luck's relevance-irrelevance
criterion described in Sec.~\ref{s}.\cite{marg} For the IM, the crossover
exponent in Eq.~(\ref{S.7}) with $\nu\!=\!1$ is $\phi_{\rm IM}\!=\!0$ whereas for the
DW, with $\nu_{\perp}\!=\!1/2$, it is $\phi_{\rm DW}\!=\!1/2$. Consequently, the
nonperiodic perturbation is marginal at the homogeneous ($R\!=\!1$) IM fixed
point whereas it is relevant for the $R\!=\!1$ DW fixed point. These statements
are in accordance with the exact results.

For the IM, the
marginal perturbation creates a line of fixed points, which is parametrized
by $R$, and the critical properties are continuously varying, even at $R\!=\!1$.
The nonperiodicity also induces a continuously varying anisotropic scaling
behavior.  However, the different varying exponents are not independent:
Knowledge of the scaling dimensions of the surface magnetization is sufficient
to completely describe the nonuniversal critical behavior studied in this work.

Considering the DW problem, here the line of fixed points is discontinuous at
$R\!=\!1$, in accordance with the relevant nature of the perturbation. For the
hierarchical models the line of fixed points is characterized
by finite coordinates and the corresponding critical behavior is of power-law
form with $R$-dependent exponents. On the other hand for the aperiodic models
(period doubling, three folding and paper folding) the line of DW fixed points
is shifted to infinity and the scaling behavior is anomalous: The
transverse fluctuations of the walk grow on a logarithmic scale.

Finally we discuss the local critical behavior at extended defects,
located either at the surface or in the bulk, which are generated by the
Fredholm sequence. In both cases two fixed points exchange there stability at a
critical value $R_c$ of the modulation amplitude $R$. This critical value
separates two regimes for the local transition: For $R\!>\!{R}_c$ the local
magnetization vanishes discontinuously at the bulk critical point while for
$R\!<\!{R}_c$ the transition is continuous. In both cases one obtains critical
exponents which vary continuously with the marginal parameter $R$.
 
It has been already noticed\cite{karevski95} that
the surface Fredholm perturbation is closely connected to the Hilhorst--van
Leeuwen model.\cite{hvl} In the same way, the bulk Fredholm defect
is connected to the Bariev model.\cite{bariev88} In these
models, the perturbation of the couplings decays as a power of the distance $l$
from the center of the defect with $\delta\lambda(l)\!=\!\alpha l^{-1}$ in the
marginal case, for the 2D IM. 

The varying exponents obtained analytically and numerically in
Ref.~\onlinecite{karevski95} for the surface Fredholm perturbation as well as
those otained via exact RG transformations in Sec.~\ref{f} for the
surface and bulk Fredholm defects can be put in correspondance with the
exponents of the Hilhorst--van Leeuwen and Bariev models with $\alpha$
replaced by $\ln R/\ln2$. Up to now, the values of the Bariev model's
exponents had been conjectured on the basis of conformal methods
using gap-exponent relations after a conformal transformation of
the inhomogeneous infinite system onto an inhomogeneous infinite
strip with periodic boundary conditions.\cite{igloi90} Our RG results for the
bulk Fredholm defect and the correspondence between both models strongly
support the validity of this procedure.

\acknowledgments

This work has been supported by the French-Hun\-ga\-rian coo\-pe\-ration program
``Balaton" (Mini\-st\`ere des Affaires Etrang\`eres--O.M.F.B.), the
Hun\-ga\-rian Na\-tio\-nal Re\-search Fund  under Grant Nos. OTKA TO 12830,
17\-485 and 23642, and the Hungarian Ministery of Edu\-ca\-tion under Grant No.
FKFP0765/1997. The Labo\-ratoire de Physique des Mat\'eriaux is Unit\'e de
Recherche Asso\-ci\'ee au C.N.R.S. No. 155.

\begin{appendix}

\section{Renormalization of the period-doubling sequence}
\label{apd}
The renormalization of the period-doubling sequence in the bulk involves the
following set of equations:
\begin{mathletters}
\label{PD.4} 
\begin{eqnarray}
&&\lambda R^{f_{4k}}\,
V(8k)-\widehat{\Lambda}\, V(8k\!+\!1)+V(8k\!+\!2)=0\; ,\label{PD.4a}\\
&&V(8k\!+1\!)-\widehat{\Lambda}\, V(8k\!+\!2)+\lambda\, V(8k\!+\!3)=0\;
,\label{PD.4b}\\ 
&&\lambda\, V(8k\!+\!2)-\widehat{\Lambda}\, V(8k\!+\!3)+V(8k\!+\!4)=0\;
,\label{PD.4c}\\  
&&V(8k\!+\!3)-\widehat{\Lambda}\, V(8k\!+\!4)+\lambda R\, V(8k\!+\!5)=0\;
,\label{PD.4d}\\ 
&&\lambda R\, V(8k\!+\!4)-\widehat{\Lambda}\, V(8k\!+\!5)+V(8k\!+\!6)=0\;
,\label{PD.4e}\\ 
&&V(8k\!+\!5)-\widehat{\Lambda}\, V(8k\!+\!6)+\lambda\, V(8k\!+\!7)=0\;
,\label{PD.4f}\\ 
&&\lambda\, V(8k\!+\!6)-\widehat{\Lambda}\, V(8k\!+\!7)+\lambda\,V(8k\!+\!8)=0\;
,\label{PD.4g}\\   
&&V(8k\!+\!7)-\widehat{\Lambda}\, V(8k\!+\!8)+\lambda R^{f_{4k\!+\!4}}\,
V(8k\!+\!9)=0\; .  
\label{PD.4h} 
\end{eqnarray}
\end{mathletters}
Among these eight equations we eliminate the six central ones,
which amounts to rescale the system by a factor $b\!=\!4$. This is accomplished
by evaluating $V(8k\!+\!2)$ and $V(8k\!+\!7)$ as functions of $V(8k\!+\!1)$ and 
$V(8k\!+\!8)$, in the linear system given by Eqs.~(\ref{PD.4}b)--(\ref{PD.4}g),
with the result
\begin{eqnarray}
&&V(8k\!+\!2)={d\widehat{\Lambda}\over c}\, V(8k\!+\!1)+{R\lambda^3\over c}\,
V(8k\!+\!8)\; ,\nonumber\\
&&V(8k\!+\!7)={R\lambda^3\over c}\, V(8k\!+\!1)+{d\widehat{\Lambda}\over c}\,
V(8k\!+\!8)\; ,\nonumber\\
&&c=\widehat{\Lambda}^2(\widehat{\Lambda}^2-\lambda^2-1)^2
-\lambda^2R^2(\widehat{\Lambda}^2-\lambda^2)^2\; ,\nonumber\\
&&d=(\widehat{\Lambda}^2-1)(\widehat{\Lambda}^2-\lambda^2-1)
-\lambda^2R^2(\widehat{\Lambda}^2-\lambda^2)\; ,
\label{PD.5}
\end{eqnarray}
Inserting these values into Eqs.~(\ref{PD.4a}) and~(\ref{PD.4h}), after
multiplication by $c/(R\lambda^3)$ we obtain
\begin{eqnarray}
&&{c\over R\lambda^2}R^{f_{4k}}\, V(8k)-\widehat{\Lambda}\,{c-d\over
R\lambda^3}\, V(8k\!+\!1)+V(8k\!+\!8)=0\, ,\nonumber\\
&&V(8k\!+\!1)-\widehat{\Lambda}\,{c-d\over R\lambda^3}\, V(8k\!+\!8)+\nonumber\\
&&\qquad\qquad\qquad\qquad +{c\over R\lambda^2}R^{f_{4k+4}}\, V(8k\!+\!9)=0\, ,
\label{PD.6}
\end{eqnarray}
which are the renormalized equations
\begin{eqnarray}
&&\lambda' R^{f_k}\, V'(2k)-\widehat{\Lambda}'\, V'(2k\!+\!1)+V'(2k\!+\!2)=0\;
,\nonumber\\ 
&&V'(2k\!+\!1)\!-\!\widehat{\Lambda}' V'(2k\!+\!2)\!+\!\lambda' R^{f_{k+1}}
V'(2k\!+\!3)=0\, , \label{PD.7}
\end{eqnarray}
after rescaling by $b\!=\!4$. Noticing that, according to Eqs.~(\ref{PD.3}),
$f_{4k}\!=\!1\!-\!{}f_{2k}\!=\!{}f_k$, $R$ remains unchanged and one obtains the
RG transformation as given in Eq.~(\ref{PD.8}).

At the surface, in terms of the reduced variables, the
same set of equations as in Eqs.~(\ref{PD.4}) with $k\!=\!0$ is obtained, except
for the two first equations which now read 
\begin{mathletters} 
\label{PD.21} 
\begin{eqnarray}
&&-\widehat{\Lambda}\, V(1)+\theta t_s\, V(2)=0\; ,\label{PD.21a}\\
&&{t_s\over\theta}\, V(1)-\widehat{\Lambda}\, V(2)+\lambda\,
V(3)=0\; .
\label{PD.21b}
\end{eqnarray}
\end{mathletters}
The auxiliary variable $\theta$ is needed to take into account the
asymmetry resulting from the renormalization after one step. In this way the
variables $\widehat{\Lambda}$, $\lambda$, $t_s$, and $\theta$ build a
closed set under renormalization. 

As above, the components $V(2)$ and $V(7)$ can be deduced from the six
central equations and read 
\begin{eqnarray}
&&V(2)={d\widehat{\Lambda}t_s\over c\theta}\, V(1)+{R\lambda^3\over
c}\, V(8)\; ,\nonumber\\ 
&&V(7)={R\lambda^3t_s\over c\theta}\,
V(1)+{d\widehat{\Lambda}\over c}\, V(8)\; ,
\label{PD.22} 
\end{eqnarray}
where $c$ and $d$ are defined in Eqs.~(\ref{PD.5}). Equations~(\ref{PD.21a})
and~(\ref{PD.4h}), after multiplication by appropriate factors, then give
\begin{eqnarray}
&&-\widehat{\Lambda}\,{c-d\over
R\lambda^3}\, V(1)+\theta t_s\,{c-d\over c-dt_s^2}\, V(8)=0\;
,\nonumber\\ 
&&{t_s\over\theta}\, V(1)-\widehat{\Lambda}\,{c-d\over R\lambda^3}\,
V(8)+{c\over R\lambda^2}\, V(9)=0\; .
\label{PD.23}
\end{eqnarray}
These equations give the renormalized forms of Eqs. (\ref{PD.21a})
and~(\ref{PD.21b}) and provide the RG recursions given in Eqs.~(\ref{PD.24}).

\section{Renormalization of the hierarchical sequence}
\label{ah}
In the bulk, the set of eigenvalue equations we consider is the following:
\begin{mathletters}
\label{H.7}
\begin{eqnarray}
&&\lambda R^n
V(2m^n)\!-\!\widehat{\Lambda}V(2m^n\!+\!1)\!+\!{}V(2m^n\!+\!2)\!=\!0\, ,
\label{H.7a}\\  
&&V(2m^n\!+\!1)\!-\!\widehat{\Lambda}V(2m^n\!+\!2)\!+\!\lambda
V(2m^n\!+\!3)\!=\!0\, , 
\label{H.7b}\\ 
&&\lambda
V(2m^n\!+\!2)\!-\!\widehat{\Lambda}V(2m^n\!+\!3)\!+\!{}V(2m^n\!+\!4)\!=\!0\, ,
\label{H.7c}\\ 
&&\qquad\qquad\qquad\qquad\vdots\hfill\nonumber\\ 
&&\lambda
V(2m^n\!+\!2m\!-\!2)\!-\!\widehat{\Lambda}\, V(2m^n\!+\!2m\!-\!1)+\nonumber\\
&&\qquad\qquad\qquad\qquad\qquad\quad +V(2m^n\!+\!2m)=0\, ,
\label{H.7d}\\ 
&&V(2m^n\!+\!2m\!-\!1)\!-\!\widehat{\Lambda}\, V(2m^n\!+\!2m)+\nonumber\\
&&\qquad\qquad\qquad\qquad+\lambda R\, V(2m^n+2m+1)=0\, .
\label{H.7e}
\end{eqnarray} 
\end{mathletters} 
Among the $2m$ equations one eliminates the
$2m\!-\!2$ central ones, which amounts to rescale the system by a factor of
$b\!=\!{m}$. Then we are left with two equations between the components
$V(2m^n)$, $V(2m^n\!+\!1)$, $V(2m^n\!+\!2m)$ and $V(2m^n\!+\!2m\!+\!1)$ of 
the form
\begin{eqnarray}
&&{\lambda R^n \over r} V(2m^n)\!-\!{\widehat{\Lambda}\!-\!{}s\over r}
V(2m^n\!+\!1)\!+\!{}V(2m^n\!+\!2m)\!=\!0\, ,\nonumber\\
&&V(2m^n\!+\!1)-{\widehat{\Lambda}\!-\!{}s\over
r}V(2m^n\!+\!2m)+\nonumber\\
&&\qquad\qquad\qquad\qquad+{\lambda R \over r} V(2m^n\!+\!2m+1)=0\; .
\label{H.8}
\end{eqnarray}
Here $r\!=\!\lambda^{m-1}/D_{2m-2}$, whereas $s\!=\!-D_{2m-3}/D_{2m-2}$, and
$D_{2m-2}$ denotes the $(2m\!-\!2)\!\times\!(2m\!-\!2)$ determinant
\begin{equation}
D_{2m-2}=\left|
{\renewcommand{\arraystretch}{.7}
\begin{array}{cccccc}
-\widehat{\Lambda}&\lambda&&&&\\
\lambda&-\widehat{\Lambda}&1&&&\\
&1&-\widehat{\Lambda}&\lambda&&\\
&&&\ddots&&\\
&&&1&-\widehat{\Lambda}&\lambda\\
&&&&\lambda&-\widehat{\Lambda}
\end{array}
}
\right|
\label{H.9}
\end{equation}
while $D_{2m-3}$ is the lower central minor of $D_{2m-2}$. Then
from Eqs.~(\ref{H.8}) we deduce the RG transformation given in
Eqs.~(\ref{H.10}).

\section{Renormalization of the three-folding sequence}
\label{atf}
The eigenvalue equations take the following form in the bulk:
\begin{mathletters}
\label{TF.4}
\begin{eqnarray}
&&R^{f_{3k}}V(6k)\!-\!\kappa\widetilde{\Lambda}\, V(6k\!+\!1)\!+\!\mu\,
V(6k\!+\!2)\!=\!0\, , 
\label{TF.4a}\\
&&\mu V(6k\!+\!1)\!-\!{\widetilde{\Lambda}\over\kappa}
V(6k\!+\!2)\!+\!{}V(6k\!+\!3)\!=\!0\, , 
\label{TF.4b}\\
&&V(6k\!+\!2)\!-\!\kappa\widetilde{\Lambda} V(6k\!+\!3)\!+\!\mu V(6k\!+\!4)
\!=0\!\, ,
\label{TF.4c}\\
&&\mu V(6k\!+\!3)\!-\!{\widetilde{\Lambda}\over\kappa} V(6k\!+\!4)\!+\!{}R
V(6k\!+\!5)\!=\!0\, , 
\label{TF.4d}\\
&&R V(6k\!+\!4)\!-\!\kappa\widetilde{\Lambda}\, V(6k\!+\!5)\!+\!\mu\,
V(6k\!+\!6)\!=\!0\, , 
\label{TF.4e}\\
&&\mu V(6k\!+\!5)-{\widetilde{\Lambda}\over\kappa}\,
V(6k\!+\!6)\!+\!{}R^{f_{3k+3}}\, V(6k\!+\!7)\!=\!0\, . 
\label{TF.4f}
\end{eqnarray}
\end{mathletters}
Equations~(\ref{TF.4}b)--(\ref{TF.4}e) can be used to write
\begin{eqnarray}
&&V(6k\!+\!2)\!=\!{c\kappa\widetilde{\Lambda}\over e\mu} V(6k\!+\!
1)\!+\!{R\mu^2\over e} V(6k+6)\, ,\nonumber\\
&&V(6k\!+\!5)\!=\!{R\mu^2\over e} V(6k\!+\!1)\!+\!{d\widetilde{\Lambda}\over
e\kappa\mu} V(6k\!+\!6)\, ,\nonumber\\
&&c\!=\!\mu^2(\widetilde{\Lambda}^2\!-\!\mu^2\!-\!{}R^2)\, ,
\quad d\!=\!\mu^2(\widetilde{\Lambda}^2\!-\!\mu^2\!-\!1)\, ,\nonumber\\
&&e\!=\!(\widetilde{\Lambda}^2\!-\!1)(\widetilde{\Lambda}^2\!-\!
R^2)\!-\!\mu^2\widetilde{\Lambda}^2\, , \label{TF.5}
\end{eqnarray}
which, inserted into Eqs.~(\ref{TF.4a}) and~(\ref{TF.4f}), lead to the
renormalized equations
\begin{eqnarray}
&&R^{f_{3k}}V(6k)\!-\!\kappa\widetilde{\Lambda}\left(1\!-\!{c\over e}\right)
V(6k\!+\!1)\!+\!{R\mu^3\over e}V(6k\!+\!6)\!=\!0\, ,\nonumber\\
&&{R\mu^3\over
e}V(6k\!+\!1)\!-\!{\widetilde{\Lambda}\over\kappa}\left(1\!-\!{d\over e}\right)
V(6k\!+\!6)+\nonumber\\ &&\qquad\qquad\qquad\qquad\qquad\quad
+R^{f_{3k+3}}V(6k\!+\!7)=0\, . 
\label{TF.6} 
\end{eqnarray}
Since $f_{3k}\!=\!{}f_k$ according to Eqs.~(\ref{TF.3}), these equations take the
form \begin{eqnarray}
&&R^{f_{k}} V'(2k)\!-\!\kappa'\widetilde{\Lambda}' V'(2k\!+\!1)
\!+\!\mu' V'(2k\!+\!2)\!=\!0\, ,\nonumber\\
&&\mu' V'(2k\!+\!1)\!-\!{\widetilde{\Lambda}'\over\kappa'} V'(2k\!+\!
2)\!+\!{}R^{f_{k+1}} V'(2k\!+\!3)=0\, , 
\label{TF.7}
\end{eqnarray}
with the renormalized variables given in Eqs.~(\ref{TF.8})

At the surface, Eqs.~(\ref{TF.4}c)--(\ref{TF.4}f) with $k\!=\!0$ have to be
supplemented by  
\begin{eqnarray}
&&-\kappa\widetilde{\Lambda}\, V(1)+\theta\mu_s\, V(2)=0\; ,\nonumber\\
&&{\mu_s\over\theta}\, V(1)-{\widetilde{\Lambda}\over\kappa}\, V(2)+V(3)=0\; .
\label{TF.14}
\end{eqnarray}
Rewriting $V(2)$ and $V(5)$ as functions of $V(1)$ and $V(6)$ in the first
and last equations of the surface block, one obtains the renormalized
equations
\begin{eqnarray}
&&-\kappa\widetilde{\Lambda}\,\left(1\!-\!{c\over e}\right) V(1)+
{\theta\mu_s\mu^2R\over e}{c\!-\!{}e\over
c(\mu_s/\mu)^2\!-\!e} V(6)\!=\!0\, ,\nonumber\\
&&{\mu_s\mu^2R\over\theta e} V(1)\!-\!{\widetilde{\Lambda}\over\kappa}
\left(1\!-\!{d\over e}\right) V(6)\!+\!{}V(7)\!=\!0\, .
\label{TF.15}
\end{eqnarray}
A comparison with Eqs.~(\ref{TF.14}) leads to the renormalized parameters given
in Eqs.~(\ref{TF.16}).

\section{Renormalization of the paper-folding sequence}
\label{apf}
The following blocks have to be considered,
\begin{mathletters}
\label{PF.4}
\begin{eqnarray}
&&R^{f_{4k}} V(8k)\!-\!\kappa_\alpha\kappa\widetilde{\Lambda} V(8k\!+\!
1)\!+\!\mu_\alpha V(8k\!+\!2)\!=\!0\, ,\label{PF.4a}\\
&&\mu_\alpha V(8k\!+\!1)\!-\!{\widetilde{\Lambda}\over\kappa} V(8k\!+\!
2)\!+\!V(8k\!+\!3)\!=\!0\, , \label{PF.4b}\\
&&V(8k\!+\!2)\!-\!\kappa\widetilde{\Lambda} V(8k\!+\!3)\!+\!\mu_\beta 
V(8k\!+\!4)\!=\!0\, , 
\label{PF.4c}\\
&&\mu_\beta V(8k\!+\!3)\!-\!{\kappa_\beta\widetilde{\Lambda}\over\kappa} 
V(8k\!+\!4)\!+\! R^{f_{4k+2}} V(8k\!+\!5)\!=\!0 ,\label{PF.4d}
\end{eqnarray}
\end{mathletters}
when the central interaction is $J$ and
\begin{mathletters}
\label{PF.5}
\begin{eqnarray}
&&R^{f_{4k+2}} V(8k\!+\!4)\!-\!\kappa_\alpha\kappa\widetilde{\Lambda}
V(8k\!+\!5)\!+\!\mu_\alpha V(8k\!+\!6)\!=\!0 ,\label{PF.5a}\\
&&\mu_\alpha V(8k\!+\!5)\!-\!{\widetilde{\Lambda}\over\kappa} V(8k\!+\!6)
\!+\!R V(8k\!+\!7)\!=\!0\, , \label{PF.5b}\\
&&R V(8k\!+\!6)\!-\!\kappa\widetilde{\Lambda} V(8k\!+\!7)\!+\!\mu_\beta
V(8k\!+\!8)\!=0\!\, , \label{PF.5c}\\
&&\mu_\beta V(8k\!+\!7)\!-\!{\kappa_\beta\widetilde{\Lambda}\over\kappa} 
V(8k\!+\!8)\!+\! R^{f_{4k+4}} V(8k\!+\!9)\!=\!0 ,\label{PF.5d}
\end{eqnarray}
\end{mathletters}
when the central interaction is $RJ$. $\mu_\alpha\!=\!{h}_\alpha/J$ and
$\mu_\beta\!=\!{h}_\beta/J$ are reduced temperaturelike parameters and
$\widetilde{\Lambda}$ the reduced eigenvalue defined before. Since in each block two
sites out of four are eliminated, lengths are rescaled by a factor $b\!=\!2$.

The two intermediate equations in Eqs.~(\ref{PF.4}) and~(\ref{PF.5}) give
\begin{eqnarray}
&&V(8k\!+\!
2)\!=\!{\kappa\mu_\alpha\widetilde{\Lambda}\over\widetilde{\Lambda}^2\!-\!1} 
V(8k\!+\!1)\!+\!
{\mu_\beta\over\widetilde{\Lambda}^2\!-\!1} V(8k\!+\!4)\, ,\nonumber\\
&&V(8k\!+\!3)\!=\!{\mu_\alpha\over\widetilde{\Lambda}^2\!-\!1} V(8k\!+\!1)\!+\!
{\kappa^{-1}\mu_\beta\widetilde{\Lambda}\over\widetilde{\Lambda}^2\!-\!1} 
V(8k\!+\!4)\, ,\nonumber\\
&&V(8k\!+\!
6)\!=\!{\kappa\mu_\alpha\widetilde{\Lambda}\over\widetilde{\Lambda}^2\!-\!{}R^2}
V(8k\!+\!5)\!+\!{\mu_\beta R\over\widetilde{\Lambda}^2\!-\!{}R^2} 
V(8k\!+\!8)\, ,\nonumber\\ 
&&V(8k\!+\!7)\!=\!{\mu_\alpha R\over\widetilde{\Lambda}^2\!-\!{}R^2}
V(8k\!+\!
5)\!+\!{\kappa^{-1}\mu_\beta\widetilde{\Lambda}\over\widetilde{\Lambda}^2\!-\!{}R^2}
V(8k\!+\!8)\, , 
\label{PF.6}
\end{eqnarray} 
which can be used in the first and
last lines of Eqs.~(\ref{PF.4}) and~(\ref{PF.5}), together with the first
relation in Eqs.~(\ref{PF.3}), to write the renormalized equations 
\begin{eqnarray} 
&&R^{f_{2k}} V'(4k)\!-\!\kappa_\alpha'\kappa'\widetilde{\Lambda}' 
V'(4k\!+\!1)\!+\!\mu_\alpha' V'(4k\!+\!2)\!=\!0 ,\nonumber\\
&&\mu_\alpha' V'(4k\!+\!1)\!-\!{\widetilde{\Lambda}'\over\kappa'} V'(4k\!+\!2)
\!+\!{}R^{f_{2k+1}} V'(4k\!+\!3)\!=\!0 ,\nonumber\\
&&R^{f_{2k+1}} V'(4k\!+\!2)\!-\!\kappa'\widetilde{\Lambda}' V'(4k\!+\!
3)\!+\!\mu_\beta' V'(4k\!+\!4)\!=\!0 ,\nonumber\\
&&\mu_\beta' V'(4k\!+\!3)\!-\!{\kappa_\beta'\widetilde{\Lambda}'\over\kappa'}
V'(4k\!+\!4)\!+\!{}R^{f_{2k+2}} V'(4k\!+\!5)\!=\!0 .
\label{PF.7}
\end{eqnarray}
Here the components of the eigenvectors transform according to
\begin{eqnarray}
&&V'(4k)\!=\!{}V(8k)\, ,\qquad V'(4k\!+\!1)\!=\!{}V(8k\!+\!1)\, ,\nonumber\\
&&V'(4k\!+\!2)\!=\!{}V(8k\!+\!4)\, ,\quad V'(4k\!+\!3)\!=\!{}V(8k\!+\!5)\, .
\label{PF.8}
\end{eqnarray}
The renormalized parameters are given in Eqs.~(\ref{PF.9}).

The surface field $h_s\!=\!{J}\zeta_s$ introduces a supplementary equation in the
surface block which now begins with
\begin{eqnarray}
&&-{\kappa_\beta\widetilde{\Lambda}\over\kappa}\, V(0)+\zeta_s V(1)=0\;
,\nonumber\\
&&\zeta_s\, V(0)-\kappa_\alpha\kappa\widetilde{\Lambda} V(1)+\theta\mu_s 
V(2)=0\; ,\nonumber\\
&&{\mu_s\over\theta}\, V(1)-{\widetilde{\Lambda}\over\kappa}\, V(2)+V(3)=0\; ,
\label{PF.13}
\end{eqnarray}
where, as before, $\mu_s\!=\!{}h_1/J$ is a temperaturelike surface variable
and $\theta$ an auxiliary variable. 

The first equation in Eqs.~(\ref{PF.13}) gives the value of $V(0)$ which can be
used in the second equation to obtain a surface block in its standard form:
\begin{eqnarray}
&&-\kappa\widetilde{\Lambda}\left(\kappa_\alpha
-{\zeta_s^2\over\kappa_\beta\widetilde{\Lambda}^2}\right) V(1)+\theta\mu_s V(2)=0\;
,\nonumber\\
&&{\mu_s\over\theta}\, V(1)-{\widetilde{\Lambda}\over\kappa}\, V(2)+V(3)=0\; .
\label{PF.14}
\end{eqnarray}
The two remaining equations are given by~(\ref{PF.4c}) and~(\ref{PF.4d}) with
$k\!=\!0$. As usual, writing $V(2)$ and $V(3)$ as functions of $V(1)$ and $V(4)$,
one obtains the renormalized equations
\begin{eqnarray}
&&-d_\alpha\kappa\widetilde{\Lambda}\,\left(1-{\zeta_s^2\over
d_s\kappa_\beta\widetilde{\Lambda}^2}\,\right) V(1)
+{\theta\mu_sd_\alpha\mu_\beta\over d_s(\widetilde{\Lambda}^2\!-\!1)}\, V(4)=0
\; ,\nonumber\\
&&{\mu_s\mu_\beta\over\theta(\widetilde{\Lambda}^2\!-\!1)}\,
V(1)-{d_\beta\widetilde{\Lambda}\over\kappa}\, V(4)+V(5)=0\; ,
\label{PF.15}
\end{eqnarray}
where $d_s\!=\!\kappa_\alpha\!-\!\mu_s^2/(\widetilde{\Lambda}^2\!-\!1)$.
When compared to Eqs.~(\ref{PF.14}) they lead to the renormalized parameters
given in Eqs.~(\ref{PF.16}).

\section{Renormalization of the Fredholm sequence}
\label{af}
We have to consider the following block of equations:
\begin{mathletters}
\label{F.3}
\begin{eqnarray}
&&\lambda R^{f_{2k}} V(4k)\!-\!\widehat{\Lambda} V(4k\!+\!1)\!+\!V(4k\!+\!
2)\!=\!0\, ,\label{F.3a}\\
&&V(4k\!+\!1)\!-\!\widehat{\Lambda} V(4k\!+\!2)\!+\!\lambda V(4k\!+\!3)\!
=\!0\, ,
\label{F.3b}\\ 
&&\lambda V(4k\!+\!2)\!-\!\widehat{\Lambda} V(4k\!+\!3)\!+\!{}V(4k\!+\!
4)\!=\!0\, ,
\label{F.3c}\\
&&V(4k\!+\!3)\!-\!\widehat{\Lambda} V(4k\!+\!4)\!+\!\lambda R^{f_{2k+2}}
V(4k\!+\!5)\!=\!0\, . 
\label{F.3d}
\end{eqnarray}
\end{mathletters}
Equations~(\ref{F.3}b) and~(\ref{F.3}c) give the eigenvector components
\begin{eqnarray}
&&V(4k\!+\!2)\!=\!{\widehat{\Lambda}\over\widehat{\Lambda}^2\!-\!\lambda^2}
V(4k\!+\!1)\!+\!{\lambda\over\widehat{\Lambda}^2\!-\!\lambda^2} V(4k\!+\!4)
 ,\nonumber\\ 
&&V(4k\!+\!3)\!=\!{\lambda\over\widehat{\Lambda}^2\!-\!\lambda^2} V(4k\!+\!1)
\!+\!{\widehat{\Lambda}\over\widehat{\Lambda}^2\!-\!\lambda^2} V(4k\!+\!4) ,
\label{F.4}
\end{eqnarray}
which can be used to rewrite the first and last equations as
\begin{eqnarray}
&&(\widehat{\Lambda}^2\!-\!\lambda^2)R^{f_{2k}}
V(4k)\!-\!{\widehat{\Lambda}\over\lambda}(\widehat{\Lambda}^2\!
-\!\lambda^2\!-\!1) V(4k\!+\!1)+\nonumber\\
&&\qquad\qquad\qquad\qquad\qquad\qquad\qquad\qquad+V(4k\!+\!4)\!=\!0\,
,\nonumber\\
&&V(4k\!+\!1)\!-\!{\widehat{\Lambda}\over\lambda}(\widehat{\Lambda}^2\!-
\!\lambda^2\!-\!1) V(4k\!+\!4)+\nonumber\\
&&\qquad\qquad\qquad\quad+(\widehat{\Lambda}^2\!-
\!\lambda^2)R^{f_{2k+2}}V(4k\!+\!5)\!=\!0\, .   \label{F.5} 
\end{eqnarray}
Since $f_{2k}\!=\!{}f_k$ and $f_{2k\!+\!2}\!=\!{}f_{k\!+\!1}$, the
renormalized equations keep their original form with the transformed parameters
given by Eqs.~(\ref{F.6}).

The surface block reads
\begin{mathletters}
\label{F.9}
\begin{eqnarray}
&&-\widehat{\Lambda}\, V(1)+t_s\, V(2)=0\; ,\label{F.9a}\\
&&t_s\, V(1)-\kappa\widehat{\Lambda}\, V(2)+\lambda R\, V(3)=0\; 
,\label{F.9b}\\
&&\lambda R\, V(2)-\widehat{\Lambda}\, V(3)+V(4)=0\; ,\label{F.9c}\\ 
&&V(3)-\widehat{\Lambda}\, V(4)+\lambda R\, V(5)=0\; .\label{F.9d}
\end{eqnarray}
\end{mathletters}
where the auxiliary variable $\kappa$ takes into account the change of the
intermediate interaction which is now $\lambda{}R$ instead of
$\lambda$ for the bulk Eqs.~(\ref{F.3}). From Eqs.~(\ref{F.9}b) and~(\ref{F.9}c)
we deduce the eigenvector components
\begin{eqnarray}
&&V(2)={t_s\widehat{\Lambda}\over\kappa\widehat{\Lambda}^2-\lambda^2R^2}\, V(1)
+{\lambda R\over\kappa\widehat{\Lambda}^2-\lambda^2R^2}\, V(4)\; ,\nonumber\\
&&V(3)={t_s\lambda R\over\kappa\widehat{\Lambda}^2-\lambda^2R^2}\, V(1) 
+{\kappa\widehat{\Lambda}\over\kappa\widehat{\Lambda}^2-\lambda^2R^2}\, V(4)\; ,
\label{F.10}
\end{eqnarray}
which are used to rewrite Eqs.~(\ref{F.9a}) and~(\ref{F.9d}) as:
\begin{eqnarray}
&&-\widehat{\Lambda}'\,\theta V(1)+t_s'\, V(4)=0\; ,\nonumber\\
&&t_s'\, \theta V(1)-\kappa'\widehat{\Lambda}'\, V(4)+\lambda'R\, V(5)=0\; .
\label{F.11}
\end{eqnarray}
In these renormalized equations $\theta$ can be interpreted as a renormalization
factor for $V(1)$ which transforms according to Eqs.~(\ref{F.12}). The
RG equations for the other parameters are given in Eqs.~(\ref{F.13}).

In the bulk, the block of eigenvalue equations to be renormalized,
corresponding to the center of the defect, is the
following: 
\begin{mathletters}
\label{F.22}
\begin{eqnarray}
&&\lambda R\, V(-2)-\widehat{\Lambda}\, V(-1)+V(0)=0\; ,\label{F.22a}\\ 
&&V(-1)-\widehat{\Lambda}\, V(0)+\lambda R\, V(1)=0\; ,\label{F.22b}\\
&&\lambda R\, V(0)-\kappa\widehat{\Lambda}\, V(1)+\tau_d\, V(2)=0\; 
,\label{F.22c}\\
&&\tau_d\, V(1)-\kappa\widehat{\Lambda}\, V(2)+\lambda R\, V(3)=0\; 
,\label{F.22d}\\
&&\lambda R\, V(2)-\widehat{\Lambda}\, V(3)+V(4)=0\; ,\label{F.22e}\\ 
&&V(3)-\widehat{\Lambda}\, V(4)+\lambda R\, V(5)=0\; .\label{F.22f}
\end{eqnarray}
\end{mathletters}
With $\lambda\!=\!0$ in Eq.~(\ref{F.22c}), the three last equations differ from
the surface Eqs.~(\ref{F.9}) only through the auxiliary factor $\kappa$
in the first one which is necessary to preserve the symmetry of the block.

Expressing $V(0)$ and~$V(3)$ in terms of $V(-1)$ and~$V(4)$, the first and last
equations of the block take the same form as the two central ones,
\begin{eqnarray}
&&\lambda' R\, V(-2)-\kappa'\widehat{\Lambda}'\, V(-1)+\tau_d'\, V(4)=0\;
,\nonumber\\ 
&&\tau_d'\, V(-1)-\kappa'\widehat{\Lambda}'\, V(4)+\lambda' R\,
V(5)=0\;  ,
\label{F.23}
\end{eqnarray}
with the renormalized local parameters given by Eqs.~(\ref{F.24}).

\end{appendix}

\end{multicols}


\begin{references}

\bibitem{periodic}M. E. Fisher, J.\  Phys.\  Soc.\  Jpn.\  Suppl.\  {\bf 26}, 
87 (1968); H. Au-Yang and B. M. McCoy, Phys.\  Rev.\  B\  {\bf 10}, 886 (1974);
P. Hoever, W. F. Wolff, and J. Zittartz, Z.\  Phys.\  B\  {\bf 44}, 129 (1981).

\bibitem{mccoy68}B. M. McCoy and T. T. Wu, Phys.\  Rev.\  {\bf 176}, 631
(1968); {\bf 188}, 982 (1969); B. M. McCoy, {\it ibid.}\  {\bf 188}, 1014
(1969); Phys.\  Rev.\  B\  {\bf 2}, 2795 (1970). 

\bibitem{fisher92}D. S. Fisher, Phys.\  Rev.\  Lett.\  {\bf 69}, 534
(1995); Phys.\ Rev.\  B\  {\bf 50}, 3799 (1994); {\bf 51}, 6411 (1995).   

\bibitem{mikheev94}L. V. Mikheev and M. E. Fisher, Phys.\  Rev.\  B\  {\bf 49},
378 (1994). 

\bibitem{fisher86}M. E. Fisher, J.\  Chem.\  Soc.\  Faraday\ 
Trans.\  {\bf 82}, 1569 (1986).  

\bibitem{randdw}B. Derrida, Phys. Scr. {\bf T38}, 6 (1991); M. Kardar, 
in {\it Fluctuating Geometries in Statistical Mechanics and Field Theory}, Les
Houches Session LXII, edited by F. David, P. Ginzparg, and J. Zinn-Justin
(Elsevier, Amsterdam, 1996), p.~1. 
 
\bibitem{shechtman84}D. Shechtman, I. Blech, D. Gratias, and J. W. Cahn,
Phys.\  Rev.\  Lett.\  {\bf 53}, 1951 (1984). 

\bibitem{multilayers}C. F. Majkrzak, J. Kwo, M. Hong, Y. Yafet, D. Gibbs, C. L.
Chien, and J. Bohr, Adv.\  Phys.\  {\bf 40}, 99 (1991).

\bibitem{review}J. M. Luck, in {\it Fundamental Problems in Statistical
Mechanics VIII}, edited by H. van Beijeren and M. H. Ernst (Elsevier,
Amsterdam, 1994), p.~127; U. Grimm and M. Baake, in {\it The
Mathematics of Aperiodic Order}, edited by R.V. Moody (Kluwer, Dordrecht, 1997),
p.~199.  

\bibitem{godreche86}C. Godr\`eche, J. M. Luck, and H. J. Orland, J.\ 
Stat.\  Phys.\ {\bf 45}, 777 (1986).

\bibitem{okabe88} Y. Okabe and K. Niizeki, J.\  Phys.\  Soc.\ 
Jpn\ {\bf 57}, 1536 (1988); E. S. S\o rensen, M. V. Jari\'c, and M. Ronchetti,
Phys.\  Rev.\  B\  {\bf 44}, 9271 (1991).

\bibitem{okabe90} Y. Okabe and K. Niizeki, J.\  Phys.\  A\  {\bf 23}, L733
(1990).

\bibitem{sakamoto89}S. Sakamoto, F. Yonezawa, K. Aoki, S. Nos\'e, and
M. Hori, J.\  Phys.\  A\ {\bf 22}, L705 (1989); C. Zhang and K. De'Bell, Phys.\ 
Rev.\  B\ {\bf 47}, 8558 (1993). 

\bibitem{langie92}G. Langie and F. Igl\'oi, J.\  Phys.\ A\ {\bf 25}, L487 (1992).

\bibitem{henley87}C. L. Henley and R. Lipowsky, Phys.\  Rev.\  Lett.\  {\bf 59},
1679 (1987); A. Garg and D. Levine, {\it ibid.}\  {\bf 59}, 1683
(1987). 

\bibitem{kogut79}J. Kogut, Rev.\  Mod.\  Phys.\  {\bf 51},
659 (1979). 
 
\bibitem{queffelec87}M. Queff\'elec, in {\it Substitutional Dynamical
Systems-Spectral Analysis}, Lecture Notes in Mathemathics, Vol. 1294,
edited by A. Dold and B. Eckmann (Springer, Berlin, 1987).
\bibitem{dumont90}J. M. Dumont, in {\it Number Theory and Physics},
Springer Proceedings in Physics, Vol. 47, edited by J. M. Luck, P.
Moussa, and  M. Waldschmidt (Springer, Berlin, 1990), p. 185.

\bibitem{tracy88a}C. A. Tracy, J.\  Stat.\  Phys.\  {\bf 51}, 481 (1988).

\bibitem{benza90}V. G. Benza, M. Kol\'ar and M. K. Ali, Phys.\  Rev.\  B\ {\bf
41}, 9578 (1990).

\bibitem{igloi88}F. Igl\'oi, J.\  Phys.\  A\  {\bf 21}, L911 (1988); M. M. Doria and I. I. Satija, Phys.\ 
Rev.\  Lett.\  {\bf 60}, 444 (1988); H. A. Ceccatto, {\it ibid.}\  {\bf
62}, 203 (1989);  Z.\  Phys.\  B\  {\bf 75}, 253 (1989); G. V. Benza,  Europhys.\  Lett.\  {\bf 8}, 321
(1989);  M. Henkel and A. Patk\'os, J.\  Phys.\  A\ {\bf 25}, 5223 (1992).

\bibitem{doria89}M. M. Doria, F. Nori, and I. I. Satija, Phys.\  Rev.\  B\  {\bf
39}, 6802 (1989); Z. Lin and R. Tao, J.\  Phys.\  A\  {\bf 25}, 2483 (1992).

\bibitem{tracy88b}C. A. Tracy, J.\  Phys.\  A\  {\bf 21}, L603 (1988).

\bibitem{luck93a}J. M. Luck, J.\  Stat.\  Phys.\  {\bf 72}, 417 (1993).

\bibitem{harris74}A. B. Harris, J.\  Phys.\  C\  {\bf 7}, 1671 (1974).

\bibitem{luck93b}J. M. Luck, Europhys.\  Lett.\  {\bf 24}, 359 (1993); F.
Igl\'oi, J.\  Phys.\  A\  {\bf 26}, L703 (1993).

\bibitem{lieb61}E. Lieb, T. Schultz, and D. Mattis, Ann.\  Phys.\ (N.Y.) {\bf
16}, 407 (1961); P. Pfeuty, Ann.\  Phys.\ (Paris) {\bf 57}, 79 (1970).

\bibitem{turban94}L. Turban, F. Igl\'oi, and B. Berche, Phys.\  Rev.\  B\  {\bf
49}, 12\, 695 (1994). 

\bibitem{igloi94}F. Igl\'oi and L. Turban, Europhys.\  Lett.\  {\bf
27}, 91 (1994).

\bibitem{berche95}B. Berche, P-E. Berche, M. Henkel, F.
Igl\'oi, P. Lajk\'o, S. Morgan, and L. Turban, J.\  Phys.\  A\  {\bf 28}, L165
(1995). 

\bibitem{igloi95}F. Igl\'oi, P. Lajk\'o, and F. Szalma, Phys.\  Rev.\ 
B\  {\bf 52}, 7159 (1995).

\bibitem{karevski95}D. Karevski, G. Pal\'agyi, and L. Turban, J.\  Phys.\  A\
{\bf 28}, 45 (1995).

\bibitem{berche96}P-E. Berche, B. Berche, and L. Turban, J.\  Phys.\  I\  {\bf
6}, 621 (1996).

\bibitem{igloi96a}F. Igl\'oi and F. Szalma, Phys.\   Rev.\   E\  {\bf 54},
1106 (1996).

\bibitem{autres}L. Turban, P. E. Berche, and B. Berche, J.\  Phys.\  A\  {\bf
27}, 6349 (1994); U. Grimm and M. Baake, J.\  Stat.\  Phys.\  {\bf 74}, 1233
(1994); F. Igl\'oi and P. Lajk\'o, J.\  Phys.\  A\ {\bf 29}, 4803 (1996); Z.\ 
Phys.\  B\  {\bf 99}, 281 (1996); J.\  Phys.\  A\ {\bf 29}, 4803 (1996); D.
Karevski and L. Turban, {\it ibid.}\ {\bf 29}, 3461 (1996).

\bibitem{rg}{\it Phase Transitions and Critical Phenomena}, 
edited by C. Domb and M. S. Green (Academic Press, London, 1976), Vol. 6. 

\bibitem{hilhorst79}H. J. Hilhorst, M. Schick, and J. M. J. van Leeuwen, 
Phys.\  Rev.\  B\  {\bf 19}, 2749 (1979). 

\bibitem{giacometti91}A. Giacometti, A. Maritan, and A. L. Stella, Int.\  J.\ 
Mod.\  Phys.\  B\  {\bf 5}, 709 (1991). %

\bibitem{griffiths69}R. B. Griffiths, Phys.\  Rev.\  Lett.\  {\bf 23},
17 (1969).

\bibitem{igloi97}F. Igl\'oi, D. Karevski, and H. Rieger (cond-mat/9707185). 

\bibitem{igloi96b}F. Igl\'oi and L. Turban, Phys.\  Rev.\  Lett.\  {\bf 77},
1206 (1996).

\bibitem{jordan28}P. Jordan and E. Wigner, Z.\  Phys.\  {\bf 47}, 631
(1928).

\bibitem{ordered}In the ordered phase $\Lambda_1\!=\!0$ and the correlation
length involves the second eigenvalue $\Lambda_2$.  

\bibitem{aniso}R. M. Hornreich, M. Luban, and S. Shtrikman, Phys.\ Rev.\ Lett.\
{\bf 35}, 1678 (1975); K. Binder and J. S. Wang, J.\ Stat.\ Phys.\ {\bf 55}, 87
(1989). 

\bibitem{pfeuty79}P. Pfeuty, Phys.\  Lett.\ {\bf 72A}, 245 (1979).

\bibitem{collet80}P. Collet and J. P. Eckmann, {\it Iterated Maps in the
Interval as Dynamical Systems}\ (Birkh\"auser, Boston, 1980).

\bibitem{sinai82}Ja. G. Sinai, Theor.\  Probab.\  Appl.\  {\bf 27}, 247 (1982).

\bibitem{peschel84}I. Peschel, Phys.\  Rev.\  B\ {\bf 30}, 6783 (1984).  

\bibitem{keirstead87}W. P. Keirstead and B. A. Huberman, Phys.\  Rev.\  A\ 
{\bf 36}, 5392 (1987).   

\bibitem{simon61}H. A. Simon and A. Ando, Econometrica {\bf 29}, 111 (1961); 
B. A. Huberman and M. Kerszberg, J.\  Phys.\  A\  {\bf 18}, L331 (1985).

\bibitem{benioff97}Such substitutions have been independently
considered in the context of quantum Turing machines: P. Benioff, Phys.\ 
Rev.\  Lett.\  {\bf 78}, 590 (1997). 

\bibitem{lin95}Z. Lin and M. Goda, Phys.\  Rev.\  B\  {\bf 51}, 6093 (1995).

\bibitem{dekking83a}M. Dekking, M. Mend\`es-France, and A. van der
Poorten, Math.\  Intelligencer\ {\bf 4}, 190 (1983).

\bibitem{dekking83b}M. Dekking, M. Mend\`es-France, and A. van der
Poorten, Math.\  Intelligencer\ {\bf 4}, 130 (1983).

\bibitem{bariev79}R. Z. Bariev, Zh.\  Eksp.\  Teor.\  Fiz.\ {\bf 77}, 
1217 (1979) [Sov. Phys. JETP {\bf 50}, 613 (1979)].

\bibitem{marg}For the hierarchical quantum IM, the relevance-irrelevance
criterion in Sec.~\protect\ref{s} does not apply directly since the
perturbation is not small (some local couplings even go to infinity). In this
case, in order to apply the criterion, one should go back to the classical
layered IM and put the hierarchical perturbation on the vertical couplings,
as described in Ref.~\onlinecite{igloi95}.  

\bibitem{hvl}H. J. Hilhorst and J. M. J. van
Leeuwen,\ Phys.\ Rev.\ Lett.\ {\bf 47}, 1188 (1981); T. W. Burkhardt, I. Guim,
H. J. Hilhorst, and J. M. J. van Leeuwen, Phys.\  Rev.\  B\ {\bf 30}, 1486
(1984); H. W. J. Bl\"ote and  H. J. Hilhorst, J.\  Phys.\  A\  {\bf 18}, 3039
(1985); T. W. Burkhardt  and F. Igl\'oi, {\it ibid.}\  {\bf 23}, L633 (1990).

\bibitem{bariev88}R. Z. Bariev, Zh.\  Eksp.\  Teor.\  Fiz.\ {\bf 94}, 
374 (1988) [Sov. Phys. JETP {\bf 67}, 2170 (1988)]. 

\bibitem{igloi90}F. Igl\'oi, B. Berche, and L. Turban, Phys.\  Rev.\  Lett.\ 
{\bf 65}, 1773 (1990).

\end{references}
\end{document}